\documentclass[prd,twocolumn,reprint,preprintnumbers,nofootinbib,superscriptaddress]{revtex4-2}
\input{header.tex}

\begin{document}

\reportnum{-2}{CERN-TH-2025-006}

\title{Advancing the phenomenology of GeV-scale axion-like particles}

\author{Maksym~Ovchynnikov}
\email{maksym.ovchynnikov@cern.ch}
\affiliation{Theoretical Physics Department, CERN, 1211 Geneva 23, Switzerland}
\author{Andrii~Zaporozhchenko}
\email{andriizaporozhchenko@knu.ua}
\affiliation{Taras Shevchenko National University of Kyiv, 64 Volodymyrs’ka str., Kyiv 01601, Ukraine}

\date{\today}

\begin{abstract}
Searches for axion-like particles (ALPs) with masses in the GeV range are a central objective of present and future Intensity Frontier experiments.  Interpreting these searches demands a reliable description of ALP production in hadronic collisions and decay. The prescription currently adopted by the community (i) depends on parameters of unphysical chiral rotation used to match gluonic ALP interactions with the interactions in terms of hadronic bound states, (ii) misdescribes the mass scaling of the ALP flux, and neglects mixing with heavy pseudoscalar resonances. We introduce a framework that treats GeV-scale ALP interactions in a chiral-rotation-invariant manner, includes their mixing with heavier excitations $\pi(1300)$, $\eta(1295)$, and $\eta(1440)$, and properly describes their production channels. When applying our description to proton beam experiments, we find that existing bounds and projected sensitivities shift by up to an order of magnitude relative to earlier estimates. We further delineate the dominant theoretical uncertainties, which originate from the still-incomplete experimental knowledge of the spectrum of pseudoscalar excitations.
\end{abstract}

\maketitle

\section{Introduction}

Axion-Like Particles (ALPs) $a$ are hypothetical pseudoscalars inspired by the QCD axion~\cite{Peccei:1977hh,Weinberg:1977ma,Wilczek:1977pj}. Their interaction Lagrangian, defined at some scale $\Lambda > \Lambda_{\text{EW}}$, often includes the terms
\begin{equation}
    \mathcal{L}_{a} \subset c_{G}\frac{\alpha_{s}}{4\pi} \frac{a}{f_{a}}G^{\mu\nu}\tilde{G}_{\mu\nu}+\frac{\partial_{\mu}a}{f_{a}}\sum_{F} c_{F}\bar{F}\gamma^{\mu}\gamma_{5}F,
    \label{eq:alp-lagrangian}
\end{equation}
where $c_{G}/f_{a},c_{F}$ are the interaction constants, $\alpha_{s}$ is the QCD running coupling, $G_{\mu\nu}, \tilde{G}^{\mu\nu}$ are the gluon field strength and its dual, and $F$ are fermion fields.

\begin{figure}[h!]
    \centering
    \includegraphics[width=0.95\linewidth]{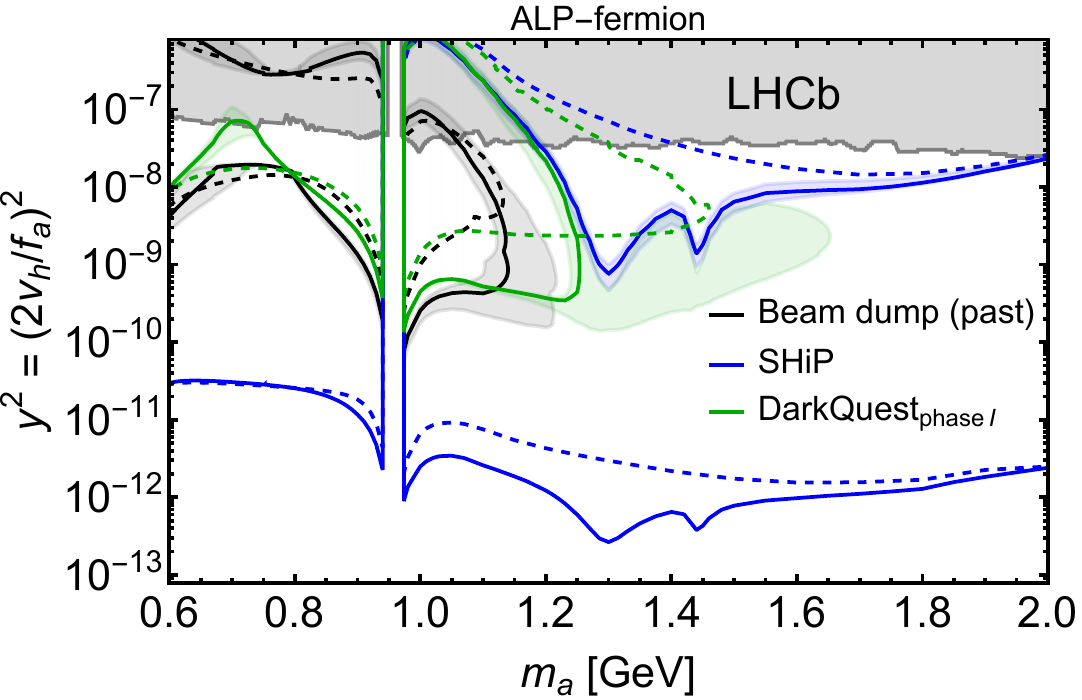}\\  \includegraphics[width=0.95\linewidth]{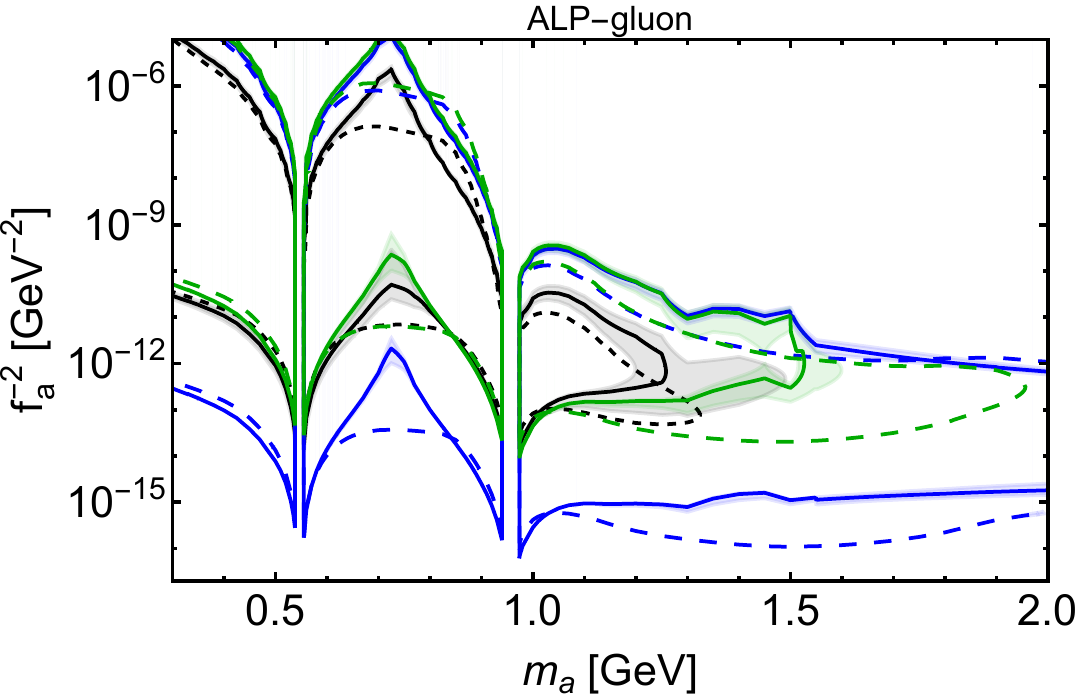}
    \caption{The parameter space of fermionic (top panel) and gluonic ALPs (bottom panel). We show the constraints/sensitivities of CHARM~\cite{CHARM:1985anb}, NuCal~\cite{Blumlein:1990ay}, BEBC~\cite{BEBCWA66:1986err}, and NA62~\cite{NA62:2023qyn} experiments (collectively marked as \texttt{Beam dumps}$_{\text{past}}$), and future DarkQuest~\cite{Batell:2020vqn} and SHiP~\cite{Aberle:2839677,SHiP:2025ows} searches. The solid lines depict the parameter space computed using the ALP phenomenology described in this work, with shaded regions estimating the theoretical uncertainties in the ALP production, while the dashed lines correspond to the approach of Refs.~\cite{Aloni:2018vki,Jerhot:2022chi,DallaValleGarcia:2023xhh}. The constraints and sensitivities have been computed using \texttt{SensCalc}~\cite{Ovchynnikov:2023cry} (see Refs.~\cite{Kyselov:2024dmi,Kyselov:2025uez} for the details of the calculation).}
    \label{fig:alp-parameter-space}
    \vspace*{-0.6cm}
\end{figure}

The ALPs with mass in the GeV range naturally appear in various extensions of the Standard Model~\cite{Essig:2013lka,Marsh:2015xka,Graham:2015ouw,Irastorza:2018dyq} and may have a relation to various cosmological phenomena, being, for example, a portal to light dark matter~\cite{Nomura:2008ru,Dolan:2014ska,Hochberg:2018rjs,Fitzpatrick:2023xks}. Because of this, they are the subject of extensive phenomenological studies~\cite{Bauer:2017ris,Aloni:2018vki,Cornella:2019uxs,Aielli:2019ivi,Bauer:2020jbp,Bauer:2021mvw,Bauer:2021wjo,Gori:2020xvq,Chakraborty:2021wda,Jerhot:2022chi,DallaValleGarcia:2023xhh,Afik:2023mhj,Bai:2024lpq}, while ALP searches are among the most important goals in the physics program of accelerator experiments. In particular, the models with $c_{G} = 1, c_{F} = 0$  and $c_{G} = 0, c_{F} = 1$ at the scale $\Lambda = 1\text{ TeV}$ (further, denoted by \emph{gluonic} and \emph{fermionic} ALPs, respectively) are considered as the benchmark models to compare the reach of various experiments~\cite{Beacham:2019nyx,Antel:2023hkf}.

The ALPs are searched for at currently running~\cite{NA62:2023qyn,Afik:2023mhj,Belle-II:2023ueh,Gorkavenko:2023nbk,NA62:2025yzs}, approved~\cite{Aberle:2839677,SHiP:2025ows,Blinov:2021say}, and proposed lifetime frontier experiments~\cite{Aielli:2019ivi,Beacham:2019nyx,Antel:2023hkf,AlemanyFernandez:2927631}. In particular, in the future SHiP experiment, ALPs may be copiously produced in collisions of a very intense proton beam with a thick target. Reconstructing their decay in a displaced decay volume allows for differentiating between the $a$ particles with different coupling patterns~\cite{Mikulenko:2023olf}. Thus, it is essential to understand the phenomenology of the GeV-scale ALPs at accelerator experiments.

The descriptions of the ALP production at proton accelerators and their decay modes from Refs.~\cite{Aloni:2018vki,Jerhot:2022chi,DallaValleGarcia:2023xhh} are incorporated in the event generators~\cite{Kling:2021fwx,Jerhot:2022chi,Ovchynnikov:2023cry} and widely adopted by the community~\cite{Antel:2023hkf}. However, they suffer from inconsistencies and miss important interactions. In this study, we address these issues, providing for the first time the description that is (i) independent of the unphysical chiral rotation eliminating the gluon coupling in Eq.~\eqref{eq:alp-lagrangian}, (ii) properly describes the ALP production modes in deep inelastic collisions, and (iii) incorporates the ALP mixing with heavy pseudoscalar excitations $\pi^{0}(1300),\eta(1295),\eta(1440)$. Using the obtained results, we refine the ALP parameter space (see Fig.~\ref{fig:alp-parameter-space}), highlighting changes in the probed couplings compared to the previous description by up to an order of magnitude.

\section{Phenomenology of ALPs: overview of state-of-the-art} 
\label{sec:state-of-the-art}
For the ALP masses $m_{a}\lesssim 1\text{ GeV}$, the description of their hadronic interactions in terms of quarks and gluons breaks down; instead, one needs to know how the $a$ particles interact with various bound states, such as mesons and baryons. For the light ALPs $m_{a}\lesssim m_{\pi}$, these interactions may be obtained by matching the operator~\eqref{eq:alp-lagrangian} and a modified Chiral Perturbation Theory (ChPT) describing the interaction of ALPs with the pseudoscalar octet $P_{8} = \pi,K,\eta$. To do this, one may first perform the chiral rotation of the quark fields:
\begin{equation}
    q \to \exp\left[-i\gamma_{5}c_{G}\hat{\kappa}_{q}\frac{a}{f_{a}}\right]q,
    \label{eq:chiral-rotation}
\end{equation}
where $\hat{\kappa}_{q}$ is a matrix satisfying the condition $\text{Tr}[\hat{\kappa}_{q}] = 1$, chosen to be diagonal. This rotation converts the gluon coupling into the derivative coupling to the quark axial-vector current~\cite{Bardeen:1986yb} (and also modifies the quark mass term), which can be translated to ChPT~\cite{Georgi:1986df,Bauer:2020jbp}. 

\begin{figure}[t!]
    \centering
    \includegraphics[width=\linewidth]{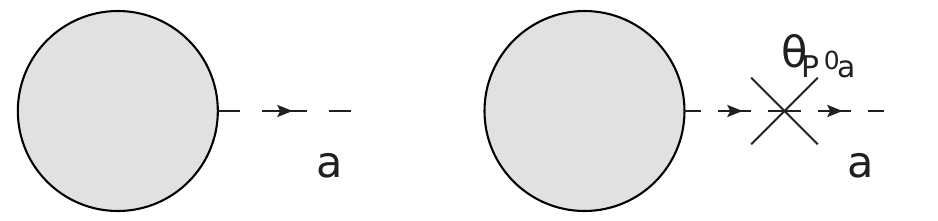}
    \caption{Practical realization of the independence of the production and decay rates of the gluonic ALP of the unphysical chiral rotation~\eqref{eq:chiral-rotation} parametrized by the matrix $\hat{\kappa}_{q}$. In the matrix element of an arbitrary process, the $\kappa_{q}$-dependent summands coming from the operators with the number of fields $\geq 3$ (the left diagram), cancel the terms originating from the quadratic Lagrangian. The latter are induced by the mixing between $a$ and pseudoscalar mesons $P^{0} = \pi^{0},\eta,\eta',\pi^{0}(1300),\dots$, and are parametrized by the mixing angles $\theta_{P^{0}a}$ (the right diagram).}
    \label{fig:mixing-direct}
\end{figure}

The resulting ChPT+ALP Lagrangian has quadratic non-diagonal terms between $a$ and $P^{0} = \pi^{0},\eta$, i.e., there is a $P^{0}-a$ mixing. The Lagrangian can be diagonalized by the linear transformation
\begin{equation}
    P^{0} \approx P^{0}_{\text{mass}} + 
    \theta_{P^{0}a}(m_{a})a + \dots,
    \label{eq:diagonalization}
\end{equation}
where the modulus of the \textit{mixing angle} $|\theta_{P^{0}a}(m_{a})|\ll 1$ everywhere except for the domain $m_{a} \simeq m_{P^{0}}$, where it gets resonantly enhanced.

The crucial next step is to perform backward matching -- ensure that the hadronic interactions at the level of mesons smoothly repeat those described by the Lagrangian~\eqref{eq:alp-lagrangian}. In particular, the hadronic decay width of ALPs calculated exclusively using the interactions with mesons must saturate the perturbative QCD prediction obtained by calculating the width into two gluons. 

\begin{table}[t!]
    \centering
    \begin{tabular}{|c|c|c|c|c|c|c|c|}
    \hline Resonance & $\eta(1295)$ & $\pi^{0}(1300)$ & $\eta(1405/1475)^{*}$ \\
        \hline Mass [GeV] & $1.294$ & $1.3$ & $1.41/1.48$ \\
        \hline Width [MeV] & $55$ & $200-600$ & $50/96$ \\ \hline
    \end{tabular}
    \caption{Properties of heavy pseudoscalar excitations we consider in this study from Particle Data Group~\cite{ParticleDataGroup:2024cfk}. Uncertainties are small for all the parameters except for the decay width of $\pi^{0}(1300)$. Asterisk in $\bm{\eta(1405/1475)}^{*}$ means that these two states may be equally well described as the single state $\eta(1440)$~\cite{Giacosa:2024epf}.}
    \label{tab:pseudoscalar-excitations}
\end{table}

The matching is performed at mass $m_{a}\gtrsim 1\text{ GeV}$. Relying solely on the pure ChPT is insufficient. First, various other mesons (such as vector, scalar, and tensor resonances) may contribute to the interactions in this mass range as intermediate states~\cite{Ilten:2018crw}. Therefore, one has to include the interactions of $a$ with these particles as well. 

Second, ALPs have mixing with the pseudoscalar singlet $\eta'$, heavier excitations 
\begin{equation}
P^{0}_{h} = \pi^{0}(1300), \quad \eta(1295), \quad \eta(1440), \quad \dots
\end{equation}
(see Table~\ref{tab:pseudoscalar-excitations}) and axial-vector mesons 
\begin{equation}
A^{0} = a_{1}^{0}, \quad f_{1}(1285), \quad f_{1}(1415)
\end{equation}
The decay widths of many of them are comparable to or smaller than the width of the $\rho$ meson. As a result, these resonances do not overlap, and their effects must be explicitly incorporated into the ALP phenomenology.

Importantly, an essential property of the resulting description of the ALP phenomenology is that it \textit{must} predict the observables that are independent of the chiral rotation $\hat{\kappa}_{q}$ from Eq.~\eqref{eq:chiral-rotation} (modulo the unambiguous contribution $\text{Tr}[\hat{\kappa}_{q}]$). This is achieved by summing the two $\kappa_{q}$-dependent contributions: from the diagonalization of the ALP-meson quadratic Lagrangian (see Fig.~\ref{fig:mixing-direct}), and from the ALP-meson operators being cubic and quartic in fields.

To the best of our knowledge, no existing study addresses all these features simultaneously. On the one hand, Refs.~\cite{Georgi:1986df,Herrera-Siklody:1996tqr,Kaiser:2000gs,Bauer:2021mvw,Bauer:2021wjo,Blinov:2021say} formulated the ALP interactions in a $\kappa_{q}$-invariant way. However, they only considered pure ChPT and the anomalous coupling of ALPs to photons. On the other hand, Refs.~\cite{Aloni:2018vki,Cheng:2021kjg,DallaValleGarcia:2023xhh} added the phenomenological interactions of ALPs with vector, scalar, and tensor mesons, and performed the matching of the ALP decay modes. However, in the case of non-zero $c_{G}$, the added interactions produced $\kappa_{q}$-dependent results: the contribution from higher-power operators was not included. In addition, they miss the mixing of ALPs with $P^{0}_{h},A^{0}$. These resonances are typically narrow and do not overlap, questioning the $\mathcal{O}(1)$ estimate of the uncertainty of the approach from Ref.~\cite{Aloni:2018vki}, which is based on the $U(3)$ symmetry restoration of the ALP representation in the mass range $1\text{ GeV}\lesssim m_{a} \lesssim 2\text{ GeV}$.

Let us now highlight another issue in the adopted ALP phenomenology description by considering the production of ALPs at proton accelerator experiments. The state-of-the-art approach from Ref.~\cite{Jerhot:2022chi} describes the ALP production in deep inelastic scatterings and decays of light mesons by the product of the fluxes of $\pi^{0},\eta,\eta'$ mesons times the corresponding squared mixing angles $|\theta_{P^{0}a}|^{2}$. It has been consequently incorporated in the event generators with LLPs~\cite{Kling:2021fwx,Antel:2023hkf,Ovchynnikov:2023cry}. Such a description is not only $\kappa_{q}$-dependent -- it also wrongly describes how the flux of ALPs depends on the ALP mass. In addition, it does not account for the theoretical uncertainty coming from different production modes.\footnote{Ref.~\cite{Blinov:2021say} has partially disentangled different production mechanisms, leaving, however, unexplored theoretical uncertainties and important production modes such as the production in fragmentation.}

\section{Our approach}
\label{sec:our-approach}
The method we develop\footnote{The approach is implemented in the Mathematica notebook from Ref.~\cite{DallaValleGarcia:2023xhh} and available on \faGithub \href{https://github.com/maksymovchynnikov/ALPs-phenomenology}{maksymovchynnikov/ALPs-phenomenology} and \href{https://doi.org/10.5281/zenodo.14616404}{\texttt{10.5281/zenodo.14616404}}.} combines the self-consistent, $\kappa_{q}$-independent approach with the data-driven description of interactions of various mesons, including the heavy pseudoscalar excitations. Its technical details are summarized in Appendix~\ref{app:ALP-interactions}, while below, we provide a summary.

We include the light pseudoscalar and scalar mesons, vector, and tensor mesons, similarly to how it is done in Refs.~\cite{Aloni:2018vki,Cheng:2021kjg}, and then extend the meson sector by adding the axial-vector $A$ and heavy pseudoscalar mesons $P_{h}$ (details are summarized in Sec.~\ref{sec:heavy-P} below).  

Here and below, we consider the three-flavor setup, incorporating the $\eta'$ meson in the pseudoscalar nonet matrix as in Refs.~\cite{Herrera-Siklody:1996tqr,Aloni:2018vki}. We calculate all the quantities in the order $\mathcal{O}(\delta)$, where $\delta \equiv (m_{d}-m_{u})/(m_{d}+m_{u})$ is the isospin parameter, and $\mathcal{O}(\epsilon)$, where $\epsilon \equiv f_{\pi}/f_{a}\ll 1$ is the ALP dimensionless coupling, with $f_{\pi} \approx 93\text{ MeV}$ being the pion decay constant. We incorporate the renormalization group flow from Ref.~\cite{Bauer:2021mvw}, describing the evolution of the fermionic coupling $c_{F}(Q,\Lambda)$ from the scale $\Lambda$, defining the Lagrangian~\eqref{eq:alp-lagrangian}, down to the scale of interest $Q\simeq 2\text{ GeV}$.

\begin{figure*}[t!]
    \centering
    \includegraphics[width=\linewidth]{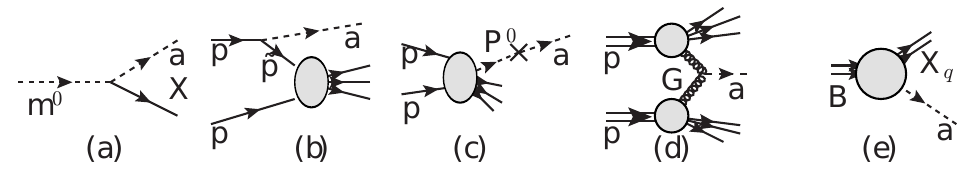}
    \caption{Diagrams of the production of ALPs at proton accelerator experiments: decays of light mesons (a), initial state radiation known as proton bremsstrahlung (b), final radiation processes in quark fragmentation (c), Drell-Yan process (d), and decays of $B$ mesons into an ALP and a hadronic state including an $s$ of $d$ quark, $X_{s/d}$ (e).}
    \label{fig:production-diagrams}
\end{figure*}

To find the mixing~\eqref{eq:diagonalization} between the ALPs and mesons, we first diagonalize the quadratic part of the ALP-meson interactions; derivations and expressions for the mixing angles $\theta_{P^{0}a}$, $\theta_{A^{0}a}$, and $\theta_{P^{0}_{h}a}$ may be found in Appendices~\ref{app:ALP-minimal-ChPT},~\ref{app:adding-Ph}.

Let us now formulate the ALP interactions with various mesons in the $\kappa_{q}$-independent fashion. We focus here solely on the gluonic coupling; treatment of the quark couplings $c_{q}$ in Eq.~\eqref{eq:alp-lagrangian} is discussed in Ref.~\cite{Bauer:2021wjo} (with Appendix~\ref{app:ALP-non-minimal} providing a summary). The effective description of Standard Model interactions of such mesons is discussed in Refs.~\cite{Fujiwara:1984mp,Guo:2011ir} (vector mesons and anomalous photon vertex),~\cite{Fariborz:1999gr} (light scalar sector),~\cite{Guo:2011ir} (tensor mesons), and~\cite{Parganlija:2012fy,Parganlija:2016yxq,Giacosa:2024epf} (vector, axial-vector, and heavy pseudoscalar excitations). On top of that, we will also need the $s\to d$ operator induced by the octet operator from~\cite{Gori:2020xvq} to describe the decays of kaons into ALPs. 

The interactions are given either in terms of the manifestly $U(3)$ covariant objects $\Sigma = \exp[2i\mathcal{P}/f_{\pi}]$ or just $\mathcal{P}$, where $\mathcal{P} = \sum_{P}t_{P}P$ is the matrix of the pseudoscalar mesons nonet. To ensure the $\kappa_{q}$-invariant results, we follow Ref.~\cite{Bauer:2021wjo} and replace the $\Sigma$ matrix with its transformed version
\begin{equation}
\Sigma \to \exp\left[ic_{G}\hat{\kappa}_{q} \frac{a}{f_{a}}\right]\cdot \Sigma \cdot \exp\left[ic_{G}\hat{\kappa}_{q} \frac{a}{f_{a}}\right],
\label{eq:sigma-transformed}
\end{equation}
and then expand the resulting Lagrangian in the powers of $\mathcal{P}$ and $a$.\footnote{A similar procedure has been adopted for the $s\to d$ and anomalous interactions with photons in~\cite{Blinov:2021say}, and for the Wess-Zumino-Witten terms in~\cite{Blinov:2021say,Bai:2024lpq}.} As for the interactions written in terms of $\mathcal{P}$, let us utilize the Baker-Campbell-Hausdorff formula: 
\begin{multline}
\exp\left[ic_{G}\hat{\kappa}_{q} \frac{a}{f_{a}}\right]\cdot \Sigma\cdot \exp\left[ic_{G}\hat{\kappa}_{q} \frac{a}{f_{a}}\right] \\ \to \exp\left[i\frac{2}{f_{\pi}}\left(\mathcal{P}(x) +\epsilon c_{G}\hat{\kappa}_{q}a \right) + \mathcal{O}(\epsilon^{2})\right]
\label{eq:Sigma-transformed}
\end{multline}
To maintain the $\kappa_{q}$ invariance, we have to replace
\begin{equation}
\mathcal{P} \to \mathcal{P} + \epsilon c_{G}\hat{\kappa}_{q}a
\label{eq:P-transformed}
\end{equation} 
The resulting Lagrangian, formulated in terms of~\eqref{eq:Sigma-transformed},~\eqref{eq:P-transformed} and after performing the diagonalization~\eqref{eq:diagonalization}, provides $\kappa_{q}$-invariant description of all ALP interactions in the $\mathcal{O}(\epsilon)$ limit, thanks to the cancellation depicted in Fig.~\ref{fig:mixing-direct}. 

\begin{figure}[t!]
    \centering
    \includegraphics[width=0.95\linewidth]{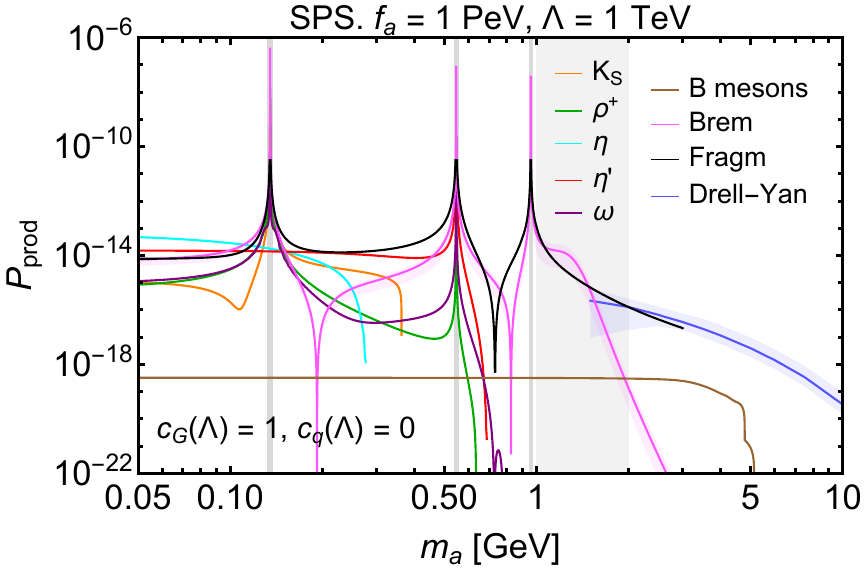} \\ \includegraphics[width=0.95\linewidth]{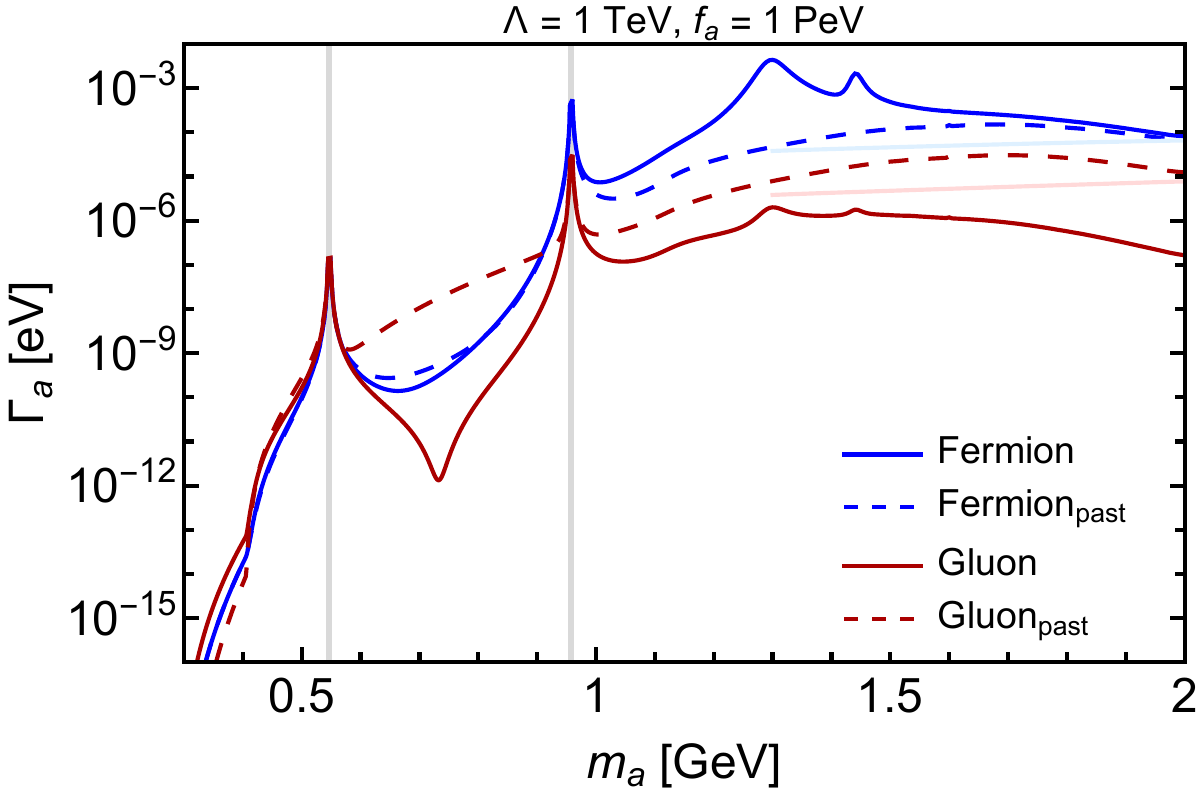}
    \caption{Phenomenology of ALPs. \textit{Top panel}: probabilities of various processes producing the gluonic ALPs (the yield of ALPs per proton-on-target) assuming the setup of a 400 GeV proton beam hitting the Molybdenum target, corresponding to the SHiP experiment~\cite{Aberle:2839677}. The bands denote theoretical uncertainties. \textit{Bottom panel}: the total hadronic decay width of the gluonic (the blue lines) and fermionic ALPs (the red lines). The dark solid lines are obtained using the approach of our work, whereas the dashed lines closely resemble the results of Ref.~\cite{Aloni:2018vki,DallaValleGarcia:2023xhh}. The light-colored lines show the perturbative QCD hadronic decay widths. The vertical bands denote the vicinities of $\pi^{0},\eta,\eta'$ masses, where the description of the ALP phenomenology based on the mixing with mesons breaks down.}
    \label{fig:production-probabilities}
\end{figure}
Importantly, using the ``rotated'' matrices~\eqref{eq:sigma-transformed},~\eqref{eq:P-transformed} not only ensures the $\kappa_{q}$ independence of the observables but also induces the additional terms with the unambiguous part $\text{Tr}[\hat{\kappa}_{q}] = 1$; therefore, it may lead to the cancellation or enhancement of the $\kappa_{q}$-independent piece.

There are two main limitations of our study. The first one is that we do not include the effects of higher-order ChPT terms on the ALP phenomenology. These effects may induce $<\mathcal{O}(1)$ corrections to $\theta_{P^{0}a}$~\cite{Alves:2017avw,Cornella:2023kjq}, but the precise calculation of their impact is limited in light of undetermined parameters of the higher order expansions. This error, however, is much smaller than the effect of the ALP interaction with heavy pseudoscalar excitations and subsequent potential uncertainty. We leave the detailed incorporation of the higher-order corrections for future work, stressing that the higher-order operators may be included within our approach similarly to other terms.

The second limitation comes from the fact that we will only account for some of the heavy resonances (see Sec.~\ref{sec:heavy-P}). We leave the inclusion of the heavier resonances $\eta(1760),\pi^{0}(1800), \eta(2225)$, as well as pseudoscalar glueballs, for future study.

\subsection{Heavy pseudoscalar excitations}
\label{sec:heavy-P}

To describe the interactions of heavy mesons, we utilize the Extended Linear Sigma Model (ELSM) framework from Ref.~\cite{Parganlija:2016yxq}, which includes the following particles that have mixing with the ALPs: neutral pseudoscalar
\begin{equation}
P_{h}^{0} = \pi^{0}(1300),\ \eta(1295), \ \eta(1440)
\end{equation}
and axial-vector mesons,
\begin{equation}
A^{0} = a_{1}^{0}, 
\ f_{1}(1285),\ f_{1}(1415)
\end{equation}
More details may be found in Appendix~\ref{app:heavy-P-AV}. The mixing with $A^{0}$\!s, however, is not resonant, which follows from the structure of the mixing operator, $\partial_{\mu}a \cdot A^{0}_{\mu}$ (see Eq.~\eqref{eq:mixing-axial-vector} of the Appendix), and their effect may be neglected.

Ref.~\cite{Parganlija:2016yxq} considered a generic expansion in terms of light pseudoscalar and heavy mesons matrices preserving the $SU_{L}(3)\otimes SU_{R}(3)$ symmetry, then dropped some of the operators for the sake of simplicity, and then fixed the expansion coefficients by requiring to recover the observable parameters such as masses and decay widths. A crucial point is that the ALP-$P^{0}_{h}$ mixing is highly sensitive to the presence of the dropped operators; we will revisit this later.

\section{ALP production} 
\label{sec:production}
We consider the following ALP production modes: the proton bremsstrahlung (approximating the initial state radiation production), quark fragmentation, 2- and 3-body decays of light mesons ($\eta, \eta', K_{S}, K^{+},\rho^{0},\omega$), the Drell-Yan process, and decays of $B$ mesons. The diagrams of these production processes are summarized in Fig.~\ref{fig:production-diagrams}.

Decays of $\rho^{0}$ and $\omega$ within the 3-flavor case are considered for the first time; we discuss their description in more detail in Appendix~\ref{app:alp-decays-mesons}.

For the description of the bremsstrahlung, we follow the quasi-real approximation from Refs.~\cite{Altarelli:1977zs,Boiarska:2019jym,Foroughi-Abari:2021zbm}. We adopt the ALP-nucleon form factor, needed for accounting for the non-point-like structure of the proton, from~\cite{Blinov:2021say}. We also incorporate the intrinsic theoretical uncertainty of the quasi-real approximation -- the hard scale defining the allowed ranges of the proton virtuality, see Refs.~\cite{Foroughi-Abari:2021zbm,Foroughi-Abari:2024xlj,Kyselov:2024dmi}. For the Drell-Yan production, we follow Ref.~\cite{DallaValleGarcia:2023xhh}; in particular, we accommodate the theoretical uncertainties by varying factorization and renormalization scales of the hard process $GG\to a$. They are sizeable because the production vertex scales with $\alpha_{s}$.

Qualitatively, the production in the fragmentation chain occurs when, because of the mixing, each $P^{0}$ appearing at the end of the fragmentation is ``replaced'' by the ALP with a tiny probability~\cite{Kyselov:2024dmi}. The probability is given in terms of ``generalized'' mixing angles, including the contributions from both the ALP-meson mixing and the multi-field operators to account for the cancellation of the unphysical dependence on the chiral rotation~\eqref{eq:chiral-rotation}, see Appendix~\ref{app:fragmentation}. This process is attractive because of a relatively small uncertainty compared to the proton bremsstrahlung for masses $m_{a}\lesssim 1\text{ GeV}$. We incorporate the production in the quark fragmentation via the mixing with light $P^{0}$\!s in \texttt{PYTHIA8}~\cite{Bierlich:2022pfr}, see Ref.~\cite{Kyselov:2025uez}.

The mixing with heavier resonances contributes to bremsstrahlung and fragmentation as well. However, including the contribution is a non-trivial task. In the bremsstrahlung case, it requires knowing the axial-vector proton elastic form factor in the time-like region, to which the heavy mesons contribute. It may be parametrized in terms of the Breit-Wigner contributions of various resonances (as is done, e.g., for dark photons in Ref.~\cite{Foroughi-Abari:2024xlj}). The coefficients in front of these contributions cannot all be fixed without accounting for the experimental data, which is very limited. As for the production in fragmentation, it requires knowing the fragmentation function into heavy hadrons. The latter must be tuned to the experimental data on the fluxes, which are lacking for heavy resonances. Therefore, our results on these modes are conservative; we leave studying these interesting questions to future work.

The summed contribution of decays of light mesons, the proton bremsstrahlung, and quark fragmentation replaces the ``flux-times-mixing-angle'' approach of Ref.~\cite{Jerhot:2022chi}, widely used in the literature~\cite{Antel:2023hkf}, see Appendix~\ref{app:production}.

The probabilities of all the production processes for the beam and target setup corresponding to the SHiP experiment for the gluonic ALPs are shown in Fig.~\ref{fig:production-probabilities} (the top panel). The uncertainty in the proton bremsstrahlung (a common feature for the other LLPs, see Ref.~\cite{Kyselov:2024dmi}) and the Drell-Yan process may reach 1-2 orders of magnitude. The production in the fragmentation chain is one of the dominant modes for gluonic ALPs, whereas decays of $B$ mesons become important for the fermionic ALPs.

\section{ALP decay modes}
\label{sec:decays}
Similar to the previous studies~\cite{Aloni:2018vki,Bauer:2021mvw,Cheng:2021kjg,DallaValleGarcia:2023xhh}, we consider exclusive ALP decay modes $a\to 2\gamma$, $3\pi$, $2\pi\gamma$, $KK\pi$, $\eta^{(')}\pi\pi$, $4\pi/2\rho^{0}$, $2\omega$, $3\eta$, $2\eta \pi$, and perturbative decays $a\to 2G,2s$. The heavy pseudoscalar excitations have the same decay modes, so the ALP mixing with them resonantly enhances the corresponding decay widths.

The total ALP hadronic decay widths for the models of fermionic and gluonic ALPs are shown in Fig.~\ref{fig:production-probabilities} (the bottom panel). We include two calculations: the one using our results, and another one utilizing the state-of-the-art approach similar to~\cite{Aloni:2018vki,DallaValleGarcia:2023xhh}, obtained in the non-covariant way (for the coupling solely to gluons) and dropping the mixing with heavy pseudoscalar mesons. To fix the $\kappa_{q}$ parameters, we adopt the common choice $\kappa_{q} = m_{q}^{-1}/\text{Tr}[m_{q}^{-1}]$, where $m_{q}$ is the quark mass matrix.

In the mass range $m_{a}\lesssim 2\text{ GeV}$, the exclusive widths differ by up to a few orders of magnitude. For the gluonic ALPs, our prediction is much smaller than the state-of-the-art results in the domain $m_{a}\gtrsim 1\text{ GeV}$. It is caused by the partial cancellation between the contributions coming from the diagonalization~\eqref{eq:diagonalization} and the rotation~\eqref{eq:sigma-transformed}, which reflects the genuine property of the gluonic interaction in Eq.~\eqref{eq:alp-lagrangian}. Because of the same cancellation, the mixing with heavy pseudoscalar mesons does not contribute significantly. These factors prevent making a match of the exclusive width with the perturbative width into gluons. The situation is different for the ALPs universally coupled to fermions, for which the heavy pseudoscalars increase the width by up to two orders of magnitude.

The predictive power of our approach (and, in principle, any ALP phenomenology study) is tied to the maturity of the effective description of the Standard Model’s heavy pseudoscalar sector encapsulated in Refs.~\cite{Parganlija:2012fy,Parganlija:2016yxq}. First, experimental information on many resonances is scarce, resulting in sizeable uncertainties on their partial widths~\cite{ParticleDataGroup:2024cfk}. Second, the internal quark composition of some of the heavy pseudoscalars remains unsettled~\cite{Giacosa:2024epf}. Finally, for simplicity, Ref.~\cite{Parganlija:2016yxq} (studying $P_{h}$ resonances) omitted several operators whose inclusion could materially alter ALP observables, see Appendix~\ref{app:open-problems}. However, including them would require a full re-analysis and refit of the operator expansion coefficients, a task
that lies beyond the scope of the present work. Nevertheless, our entire framework is implemented in a public \textsc{Mathematica} notebook, so additional operators can be incorporated straightforwardly. Continued progress in light-meson spectroscopy -- through both new data and theoretical tools such as the ELSM and the unitarization approach~\cite{ParticleDataGroup:2024cfk} -- is therefore essential for sharpening ALP phenomenology predictions.

\section{Conclusions} 
We have systematically explored the phenomenology of the ALPs in the GeV mass range at accelerator experiments. Our study has two main outcomes. First, the approach we developed (Sec.~\ref{sec:our-approach}) is simultaneously free from unphysical chiral rotation, Eq.~\eqref{eq:chiral-rotation}, and includes the ALP mixing with heavy meson resonances such as heavy pseudoscalar mesons -- the points that were missing in the previous studies, see Sec.~\ref{sec:state-of-the-art}. The heavy mesons have been incorporated using the framework of the Extended Linear Sigma Model, see Sec.~\ref{sec:heavy-P}. Incorporating these two features changes the production probabilities of ALPs by up to an order of magnitude and their decay widths by up to two orders of magnitude, see Fig.~\ref{fig:production-probabilities}, especially in the mass range $1\text{ GeV}\lesssim m_{a} \lesssim 2\text{ GeV}$. 

However, the properties of the heavy resonances are not well understood, see a discussion in Secs.~\ref{sec:production},~\ref{sec:decays}. This ignorance translates into the unavoidable limitations of understanding the ALP interactions. In light of this, our framework may serve as the new state-of-the-art study of the ALP phenomenology, with the possibility of easily incorporating new interactions once the maturity of the meson spectroscopy improves. 

The revised ALP phenomenology has a significant impact on the parameter space of ALPs, affecting the scientific program of currently running and future experiments (see Fig.~\ref{fig:alp-parameter-space}).

\section*{Acknowledgements} 

The authors thank Francesco Giacosa for useful discussions on the extended linear sigma model.

\onecolumngrid 

\appendix

\newpage

\section*{Appendix}

In this Appendix, we describe the ALP interactions with various mesons and nucleons following the self-consistent approach presented in the paper. It is organized as follows. 

Sec.~\ref{app:ALP-interactions} is devoted to the addition of ALPs to the Chiral Perturbation Theory (ChPT), describing the light pseudoscalar meson nonet, and incorporating interactions with various other mesons. There, Subsec.~\ref{app:minimal-ChPT} describes ChPT. In Subsec.~\ref{app:ALP-minimal-ChPT}, we add ALPs, define the procedure of the diagonalization of the quadratic part of the ALP-meson Lagrangian, and calculate the mixing angles. Subsec.~\ref{app:ALP-non-minimal} outlines our strategy to incorporate other mesons in the interaction Lagrangian. In Subsec.~\ref{app:adding-SVT}, we discuss the addition of light scalar, vector, and tensor mesons. Subsec.~\ref{app:nucleons} describes the ALP-nucleon interaction. Subsec.~\ref{app:heavy-P-AV} is devoted to the description of the ALP interaction with axial-vector and heavy pseudoscalar mesons. Subsec.~\ref{app:cross-checks} is devoted to cross-checks of our approach. Finally, Subsec.~\ref{app:open-problems} highlights the limitations of our approach, relating its maturity to the incomplete knowledge of meson spectroscopy in the mass range $M<2\text{ GeV}$.

Sec.~\ref{app:production} uses the results of the previous section to describe different ALP production mechanisms, including decays of light mesons and the production in the quark fragmentation. There, we also outline the problem of the widely adopted description~\cite{Jerhot:2022chi}, approximating the flux of ALPs by the flux of mesons $P^{0}$ times the squared mixing angle $|\theta_{P^{0}a}|^{2}$.

Most of the results of these sections are obtained (unless stated otherwise) using our \textsc{Mathematica} notebook, where we carefully implement the ALP phenomenology.\footnote{Available on \faGithub \href{https://github.com/maksymovchynnikov/ALPs-phenomenology}{maksymovchynnikov/ALPs-phenomenology} and \href{https://doi.org/10.5281/zenodo.14616404}{\texttt{10.5281/zenodo.14616404}}.}

\section{Meson spectroscopy and ALP interactions}
\label{app:ALP-interactions}
\subsection{Minimal ChPT}
\label{app:minimal-ChPT}
The minimal ChPT Lagrangian we will use in our studies is
\begin{equation}
    \mathcal{L}_{\text{ChPT,min}} = \frac{f_{\pi}^{2}}{2}B_{0}\text{Tr}\left[\Sigma \hat{m}^{\dagger}_{q}+\hat{m}_{q}\Sigma^{\dagger}\right] +\mathcal{L}_{\text{anomaly}}+\frac{f_{\pi}^{2}}{4}\text{Tr}\left[ D_{\mu}\Sigma D^{\mu}
    \Sigma^{\dagger}\right]
    \label{eq:pure-ChPT}
\end{equation}
Here:
\begin{itemize}
\item $B_{0} = m_{\pi^{0}}^{2}/(m_{u}+m_{d})$ and $f_{\pi} = 93\text{ MeV}$ is the pion decay constant.
\item $\Sigma$ is the matrix of the pseudoscalar mesons:
\begin{equation}
    \Sigma = \exp\left[ \frac{2i\mathcal{P}}{f_{\pi}}\right], \quad \mathcal{P} = \frac{1}{\sqrt{2}}
    \begin{pmatrix}
        \frac{\pi^{0}}{\sqrt{2}}+\frac{\eta}{\sqrt{3}}+\frac{\eta'}{\sqrt{6}} & 
    \pi^{+}& K^{+} \\ \pi^{-} & -\frac{\pi^{0}}{\sqrt{2}}+\frac{\eta}{\sqrt{3}}+\frac{\eta'}{\sqrt{6}}  & K^{0} \\ K^{-} & \bar{K}^{0} & -\frac{\eta}{\sqrt{3}} +2\frac{\eta'}{\sqrt{6}},
    \end{pmatrix} 
    \label{eq:sigma-matrix}
\end{equation}
while $D_{\mu}\Sigma = \partial_{\mu}\Sigma + ieA_{\mu}\Sigma$ is the covariant derivative. Here, we have fixed the $\eta-\eta'$ mixing angle as $\theta_{\eta\eta'} = \arcsin(-1/3)$, which provides a reasonable agreement with the experimental data while allowing us to provide the results in a simple analytic form~\cite{Aloni:2018vki}. 
\item $\hat{m}_{q} = \text{diag}(m_{u},m_{d},m_{s})$ is the matrix of quark masses. 
\item $\mathcal{L}_{\text{anomaly}}$ is the QCD anomaly-breaking term of the $U_{A}(1)$ symmetry:
\begin{equation}
    \mathcal{L}_{\text{anomaly}}=-m_{0}^{2}\eta_{0}^{2}/2,
    \label{eq:anomalous}
\end{equation}
with $\eta_{0}=\cos(\theta_{\eta \eta'})\eta' -\sin(\theta_{\eta\eta'})\eta$ being its Goldstone. The coefficient $m_{0}$ is fixed in a way such that after summing the ChPT mass term (coming from the first summand in Eq.~\eqref{eq:pure-ChPT}) and the $m_{0}$ term, there is no $\eta-\eta'$ mixing for the given $\theta_{\eta\eta'}$:
\begin{equation}
    m_{0}^{2} = \frac{3}{2}\frac{m_{\pi^{0}}^{2}(2m_{s}-m_{u}-m_{d})}{m_{u}+m_{d}}
\end{equation}
\end{itemize}

The inclusion of $\eta'$ as in~\eqref{eq:sigma-matrix} and the anomalous mass term~\eqref{eq:anomalous} are consistent with large-$N_{c}$ arguments and $SU(3)_{L}\otimes SU(3)_{R}$ symmetry~\cite{Witten:1980sp}. In particular, keeping $N_{c}$ arbitrary and expressing $\eta_{0}$ in terms of $\Sigma$, the anomalous mass term is proportional to $-1/N_{c} \ln(\text{det}[\Sigma])^{2} \sim i/N_{c}\text{Tr}[\mathcal{P}]^{2}$, which is demanded by the $1/N_{c}$ counting rules.

Calculating the masses of $\eta$ and $\eta'$, $\pi^{0}$, and using the explicit form the isospin breaking parameter $\delta = (m_{d}-m_{u})/(m_{d}+m_{u})$, one may get the following consistency relations between quark and meson masses which we will use below:
\begin{equation}
    m_{u} = m_{d}\frac{1-\delta}{1+\delta}, \quad m_{d} = (1+\delta)\frac{m_{\pi^{0}}^{2}}{2m_{\eta}^{2}-m_{\pi^{0}}^{2}}m_{s}, \quad m_{\eta'}^{2} =4m_{\eta}^{2}-3m_{\pi^{0}}^{2}, \quad m_{K^{+}}^{2} = m_{\eta}^{2}-\frac{\delta m_{\pi^{0}}^{2}}{2}
    \label{eq:masses-relations}
\end{equation}

Here and below, we will only keep the terms up to quartic in meson fields. For these purposes, for example, in $\text{Tr}[D_{\mu}\Sigma D^{\mu}\Sigma]$, we have to expand $\Sigma = \sum_{n = 0}^{4}\frac{1}{n!}\left(\frac{2i\mathcal{P}}{f_{\pi}}\right)^{n}$ and then drop all the terms with the dimensionality higher than 4.

\subsection{Adding ALPs to minimal ChPT}
\label{app:ALP-minimal-ChPT}

Let us start with the interaction Lagrangian
\begin{equation}
    \mathcal{L}_{a} = c_{G}\frac{a}{f_{a}}\frac{\alpha_{s}}{4\pi}G_{\mu\nu}^{a}\tilde{G}^{\mu\nu,a}+\frac{\partial_{\mu}a}{f_{a}}\sum_{q}c_{q}\bar{q}\gamma^{\mu}\gamma_{5}q
    \label{eq:Lagrangian-ALP}
\end{equation}
Similarly to the studies~\cite{Bauer:2020jbp,Bauer:2021mvw}, we assume that this Lagrangian is defined at some scale $\Lambda\gg \Lambda_{\text{electroweak}}$. The scales of interest are $\Lambda_{Q} \simeq m_{a}$, and the Lagrangian~\eqref{eq:Lagrangian-ALP} non-trivially evolves down to this scale: the flow induces the couplings to quarks $c_{q}(\Lambda_{Q})\neq c_{q}(\Lambda)$. In particular, even if $c_{q}(\Lambda) = 0$, we end up with non-zero $c_{q}(\Lambda_{Q})$; for the choice $\Lambda = 1\text{ TeV}$ commonly used in the literature, $c_{q} \simeq 10^{-2}$~\cite{DallaValleGarcia:2023xhh,Bauer:2021mvw}. 

The analog of the Lagrangian~\eqref{eq:Lagrangian-ALP} at the scale $\Lambda_{Q}$ can be translated to the effective Lagrangian in terms of mesonic degrees of freedom after performing the rotation
\begin{equation}
q\to \exp\left[-\frac{ic_{G}a}{f_{a}}\hat{\kappa}_{q}\gamma_{5}\right]q, \quad \text{Tr}[\hat{\kappa}_{q}] = 1
\label{eq:rotation}
\end{equation}
eliminating the direct coupling to gluons. We parametrize the $\kappa_{q}$ matrix by 
\begin{equation}
\hat{\kappa}_{q} = \text{diag}(\kappa_{u},\kappa_{d},1-\kappa_{u}-\kappa_{d})
\end{equation}
The corresponding Lagrangian in terms of the mesons is~\cite{Aloni:2018vki,Blinov:2021say,Bauer:2020jbp}
\begin{multline}
    \mathcal{L}_{\text{ChPT,min}} =\frac{1}{2}(\partial_{\mu}a)^{2}-\frac{m_{a}^{2}}{2}a^{2}+ \frac{f_{\pi}^{2}}{2}B_{0}\text{Tr}\left[\Sigma \hat{m}^{\dagger}_{q}+\hat{m}_{q}\Sigma^{\dagger}\right] +\mathcal{L}_{\text{anomaly}}+\frac{f_{\pi}^{2}}{4}\text{Tr}\left[ D_{\mu}\Sigma D^{\mu}
    \Sigma^{\dagger}\right] \\ +\frac{f_{\pi}^{2}}{2}\frac{\partial_{\mu}a}{f_{a}}\text{Tr}[(\hat{c}_{q}+c_{G}\hat{\kappa}_{q})(\Sigma D^{\mu}\Sigma^{\dagger}-\Sigma^{\dagger}D^{\mu}\Sigma)] \; 
    \label{eq:lagr-chpt-supplemental}
\end{multline}
Here: 
\begin{itemize}
\item $\hat{c}_{q} = \text{diag}(c_{u},c_{d},c_{s})$ is the direct ALP coupling to quarks. To shorten the expressions, below, we set them to zero, although keeping in all numerical calculations.
\item $\hat{m}_{q}$ becomes modified by the chiral transformation:
\begin{equation}
\hat{m}_{q} =  \exp\left[-ic_{G}\frac{a}{f_{a}}\hat{\kappa}_{q}\right]m_{q}\exp\left[-ic_{G}\frac{a}{f_{a}}\hat{\kappa}_{q}\right] \; ,
\end{equation}
\item $\mathcal{L}_{\text{anomaly}}$ is still given by Eq.~\eqref{eq:anomalous}, given that the Lagrangian~\eqref{eq:lagr-chpt-supplemental} is written already after performing the chiral rotation~\eqref{eq:rotation}. Before the rotation, it would have been given in terms of $(2c_{G}a/f_{a} - \text{Tr}[2\mathcal{P}/f_{\pi}])^{2}$ (Eq. (11) from~\cite{Witten:1980sp}), given that the ALP serves as a spurion $\theta(x)$ of the $U_{A}(1)$ transformation.
\end{itemize}
The part of the Lagrangian~\eqref{eq:lagr-chpt-supplemental} quadratic in the ALP field $a$ and flavorless pseudoscalar mesons $P^{0} = \eta,\eta',\pi^{0}$ has the form 
\begin{equation}
\mathcal{L}_{P^{0}a, \text{quad}} = \frac{1}{2}K_{ij}\partial_{\mu}X_{i}\partial^{\mu}X_{j} - \frac{1}{2}M_{ij}X_{i}X_{j},
\end{equation}
where $X = (a, \pi^{0}, \eta, \eta')$. The mass and kinetic matrices are
\begin{equation}
\hat{K} = \left(
\begin{array}{cccc}
 1 & \epsilon  c_G \left(\kappa _u-\kappa _d\right) & \sqrt{\frac{2}{3}} \epsilon  c_G \left(2 \kappa _d+2 \kappa _u-1\right) & -\frac{\epsilon  c_G \left(\kappa _d+\kappa _u-2\right)}{\sqrt{3}} \\
 \epsilon  c_G \left(\kappa _u-\kappa _d\right) & 1 & 0 & 0 \\
 \sqrt{\frac{2}{3}} \epsilon  c_G \left(2 \kappa _d+2 \kappa _u-1\right) & 0 & 1 & 0 \\
 -\frac{\epsilon  c_G \left(\kappa _d+\kappa _u-2\right)}{\sqrt{3}} & 0 & 0 & 1 \\
\end{array}
\right),
\label{eq:kinetic-matrix}
\end{equation}

{
\begin{equation}
 \hat{M} = \left(
\begin{array}{cccc}
 m_a^2 & M_{a\pi^{0}} & M_{a\eta} & M_{a\eta'} \\
M_{a\pi^{0}} & m_{\pi }^2 & -\sqrt{\frac{2}{3}} \delta  m_{\pi^{0}}^2 &
   -\frac{\delta  m_{\pi^{0}}^2}{\sqrt{3}} \\
M_{a\eta} & -\sqrt{\frac{2}{3}} \delta  m_{\pi^{0}}^2 & m_{\eta }^2 & 0 \\
 M_{a\eta'} & -\frac{\delta  m_{\pi^{0}}^2}{\sqrt{3}} & 0 & m_{\eta '}^2 \\
\end{array}
\right),
\label{eq:mass-matrix}
\end{equation}
where 
\begin{align}
M_{a\pi^{0}} &= -\epsilon  c_G m_{\pi^{0}}^2 \left((\delta +1) \kappa _d+(\delta -1) \kappa _u\right), \\ M_{a\eta} &= \sqrt{\frac{2}{3}} \epsilon  c_G \left(m_{\pi^{0}}^2 \left(\delta  \kappa _d-\delta  \kappa _u+1\right)+2 m_{\eta }^2 \left(\kappa _d+\kappa _u-1\right)\right), \\ M_{a\eta'} &= \frac{\epsilon  c_G
   \left(m_{\pi^{0}}^2 \left((\delta +3) \kappa _d-(\delta -3) \kappa _u-2\right)-4 m_{\eta }^2 \left(\kappa _d+\kappa
   _u-1\right)\right)}{\sqrt{3}}
\end{align}
For brevity, we have set here $c_{q} = 0$. However, including the quark couplings would be straightforward. In Eq.~\eqref{eq:mass-matrix}, we used the relation for the $u$ and $d$ quark masses in terms of $\delta$ and $m_{s}$, see Eq.~\eqref{eq:masses-relations}, a priori assuming that the shift in the meson masses due to ALPs is negligibly small (which is true as far as $f_{\pi}/f_{a}\ll 1$). 

A generic transformation simultaneously diagonalizing these matrices in the $\mathcal{O}(\delta)$ order is~\cite{Aloni:2018vki}
\begin{align}
    a &= a_{\text{phys}} -\sum_{P^{0} = \pi^{0},\eta,\eta'}h(P^{0},m_{P^{0}})P^{0}_{\text{phys}}, \\ P^{0} &=  P^{0}_{\text{phys}} -\sum_{P^{0'}\neq P^{0}} \frac{M_{P^{0}P^{0'}}}{m_{P^{0}}^{2} - m_{P^{0'}}^{2}} + h(P^{0},m_{a})a_{\text{phys}},
    \label{eq:diagonalization-supplemental}
\end{align}
where
\begin{equation}
    h(P^{0},m_{X}) = \frac{1}{m_{a}^{2}-m_{P^{0}}^{2}}\left[ M_{aP^{0}}-m_{X}^{2}K_{aP^{0}} + \sum_{P^{0'}\neq P^{0}}M_{P^{0}P^{0'}}\frac{M_{aP^{0'}}-m_{X}^{2}K_{aP^{0'}}}{m_{X}^{2}-m_{P^{0'}}^{2}}\right]
\end{equation}
Below, we will drop the index ``phys''.

Introducing the parameter $\epsilon = f_{\pi}/f_{a}$, the mixing angles (the coefficients in the expansion of $P^{0}$ in front of $a$) become
\begin{align}
    \theta_{\pi^{0}a} &= \frac{\frac{1}{3}\epsilon c_G \delta m_{\pi^{0}}^2\left(\frac{\left(2 m_a^2-m_{\eta '}^2-m_{\pi^{0} }^2\right)}{\left(m_a^2-m_{\eta '}^2\right)}-\frac{2\left(m_a^2-2 m_{\eta }^2+m_{\pi^{0} }^2\right)}{\left(m_a^2-m_{\eta }^2\right)}\right)}{m_a^2-m_{\pi^{0}
   }^2}+\epsilon  c_G \left(\kappa _d-\kappa _u\right), \label{eq:mixing-pi0-supplemental}\\ \theta_{\eta a} &= \frac{\sqrt{\frac{2}{3}} \epsilon c_G \left( m_a^2+m_{\pi^{0} }^2-2 m_{\eta }^2\right)}{m_a^2-m_{\eta
   }^2}-2 \sqrt{\frac{2}{3}} \epsilon  c_G \left(\kappa _d+\kappa _u\right), \label{eq:mixing-eta-supplemental}\\ \theta_{\eta'a} &= -\frac{\epsilon  c_G \left(2 m_a^2-m_{\eta '}^2-m_{\pi^{0} }^2\right)}{\sqrt{3}
   \left(m_a^2-m_{\eta '}^2\right)}+\frac{\epsilon  c_G \left(\kappa _d+\kappa _u\right)}{\sqrt{3}}
   \label{eq:mixing-etapr-supplemental}
\end{align}
If utilizing the relations~\eqref{eq:masses-relations}, the quadratic part of the Lagrangian becomes diagonalized up to terms $\mathcal{O}(\delta)$ inclusive. We note the presence of the pole terms from $\eta,\eta'$ in the $\pi^{0}a$ mixing and, vice versa, $\theta_{\eta a/\eta'a}$ include pole from $\pi^{0}$. It is caused by the $\pi^{0}-\eta/\eta'$ mixing induced by the mass matrix~\eqref{eq:mass-matrix}.

The resonant enhancement is present independently of the order of the chiral expansion, as it originates from a phenomenological quadratic Lagrangian between ALPs and $P$. The case $m_{a}\to m_{P}$ corresponds to the maximal mixing between the ALP and the $P$ states. Besides the fact that such a large mixing would require a fine-tuning of the mass parameters that is rather implausible, it would modify the properties of the $P$ in a way that is incompatible with experimental findings (e.g., with the measured rate of the $\pi^{0}\to \gamma \gamma$). In practice, we exclude the vicinity of the masses of light excitations $P^{0}$ from our analysis (see the vertical bands in Figs.~\ref{fig:alp-parameter-space},~\ref{fig:production-probabilities}).

\subsection{Interactions beyond minimal ChPT}
\label{app:ALP-non-minimal}
Let us now outline our strategy to add other interactions with ALPs besides the ``minimal ChPT''~\eqref{eq:lagr-chpt-supplemental}. The interactions are typically constructed in terms of $\Sigma, \mathcal{P}$, and their derivatives. For the purely gluonic coupling $c_{G}\neq 0, c_{q} = 0$, as discussed in the main text, we replace them under the quark chiral rotation~\eqref{eq:rotation}:
\begin{align}
\Sigma &\to \mathcal{S}\,\Sigma \,\mathcal{S}, \quad \partial_{\mu}\Sigma \to \mathcal{S}\,\left(\partial_{\mu}\Sigma +ic_{G}\frac{\partial_{\mu}a}{f_{a}}\{\hat{\kappa}_{q},\Sigma\}\right) \mathcal{S}
\label{eq:sigma-transformed-supplemental}
\\
\mathcal{P} &\to \mathcal{P} + \epsilon c_{G}\hat{\kappa}_{q}a, \quad \partial_{\mu}\mathcal{P}\to \partial_{\mu}\mathcal{P} +\epsilon c_{G}\hat{\kappa}_{q}\partial_{\mu}a
\label{eq:P-transformed-supplemental}
\end{align}
where $\mathcal{S} = \exp\left[ic_{G}\hat{\kappa}_{q} \frac{a}{f_{a}}\right]$. $\kappa_{q}$-invariance is then reached by subsequently inserting linear transformation~\eqref{eq:diagonalization-supplemental} (which modifies $\Sigma, \mathcal{P}$ matrices inside Eqs.~\eqref{eq:sigma-transformed-supplemental},~\eqref{eq:P-transformed-supplemental}), and keeping only $\mathcal{O}(\epsilon,\delta, \epsilon\cdot \delta)$ terms.

Next, let us briefly comment on the treatment of the interactions of the ALPs with the quark coupling $c_{q}\neq 0, c_{G} \neq 0$. Applying the idea of Ref.~\cite{Bauer:2021wjo}, we treat the ALP derivative in the quark coupling as a background axial-vector spurion. Hence, the quark coupling modifies Eqs.~\eqref{eq:sigma-transformed-supplemental},~\eqref{eq:P-transformed-supplemental} only in case of derivatives:
\begin{align}
    \Sigma &\to \mathcal{S}\,\Sigma \,\mathcal{S}, \quad \partial_{\mu}\Sigma \to \mathcal{S}\left(\partial_{\mu}\Sigma + i\frac{\partial_{\mu}a}{f_{a}}\{\hat{c}_{q}+c_{G}\hat{\kappa}_{q},\Sigma\} \right) \mathcal{S},     \label{eq:replacement-cq-Sigma}\\
    \mathcal{P} &\to \mathcal{P} + \epsilon c_{G}\hat{\kappa}_{q}a, \quad \partial_{\mu}\mathcal{P}\to \partial_{\mu}\mathcal{P} +\epsilon (c_{G}\hat{\kappa}_{q} + \hat{c}_{q})\partial_{\mu}a
    \label{eq:replacement-cq-P}
\end{align}
There are two exceptions to $P$ transformation~\eqref{eq:replacement-cq-P}. The first one is interactions with vector mesons and Wess-Zumino-Witten terms (Sec.~\ref{app:vector}), which effectively look like if they contain the $\mathcal{P}$ objects (see Eq.~\eqref{eq:lagr-vector}) but are genuinely formulated in terms of the derivatives $\alpha_{L/R} = (D\xi_{L/R})\xi_{L/R}^{\dagger}$~\cite{Fujiwara:1984mp,Bai:2024lpq}, where $D$ is the covariant derivative and $\Sigma = \xi^{\dagger}_{L}\xi_{R}$. In these cases, instead of utilizing~\eqref{eq:replacement-cq-P}, we replace
\begin{equation}
\mathcal{P} \to \mathcal{P} + \epsilon (c_{G}\hat{\kappa}_{q}+\hat{c}_{q})a 
\label{eq:replacement-cq-P-non-derivative}
\end{equation}
The second one is in treating the weak $s\to d$ transitions, which we elaborate on in Sec.~\ref{app:adding-s-to-d}.

The results may be obtained using the accompanying \textsc{Mathematica} notebook.

Constructing the Lagrangian behind Eq.~\eqref{eq:lagr-chpt-supplemental} using the recipes~\eqref{eq:sigma-transformed-supplemental}-\eqref{eq:replacement-cq-P-non-derivative}, we are ready to calculate the ALP production and decay rates. To do this, we first compute the matrix element of the given process and ensure that it is $\kappa_{q}$-independent. Then, following Ref.~\cite{Aloni:2018vki}, we adopt the phenomenological suppression factors $F(m)$ for the whole matrix element, accounting for the QCD sum rules. We consider the limit $\mathcal{O}(\delta)$ everywhere except for the process $a\to 3\pi$, for which the squared matrix element is already $\propto \delta^{2}$.

Below, we describe how we add various excitations: lightest scalar, vector, tensor, axial-vector, and heavy pseudoscalar. 

Regarding the sector of scalar, vector, and tensor mesons (Sec.~\ref{app:adding-SVT} for the detailed description), we mainly follow Refs.~\cite{Cheng:2021kjg,DallaValleGarcia:2023xhh}). They slightly refine the treatment of these interactions in Ref.~\cite{Aloni:2018vki} but study only the ALPs without the coupling to gluons. The mentioned refinements include: imposing unitarity constraints that restrict certain contributions to ALP decays; including the proper polarization sum of vector and tensor mesons when treating them as intermediate states; and the coupling of the tensor mesons to the stress-energy tensor in terms of pseudoscalar mesons rather than just the minimal coupling via the metrics. However, there is an important difference between our present analysis and Refs.~\cite{Cheng:2021kjg,DallaValleGarcia:2023xhh} concerning the sector of scalar mesons in our present analysis; it is discussed in Sec.~\ref{app:scalar}.

The addition of axial-vector and heavy pseudoscalar mesons is the new result that we discuss in Sec.~\ref{app:heavy-P-AV}. It is based on the so-called Extended Linear Sigma Model from Refs.~\cite{Parganlija:2012fy,Parganlija:2016yxq,Giacosa:2024epf}.

\subsection{Adding scalar, vector, and tensor resonances}
\label{app:adding-SVT}

\subsubsection{Scalar mesons}
\label{app:scalar}

We add the interactions with the lightest scalar mesons following Refs.~\cite{Black:1998wt,Fariborz:1999gr}, which introduced $SU(3)$ invariant description in terms of the matrices $\Sigma$ and the scalar nonet matrix
\begin{equation}
    \mathcal{S} = \left(
\begin{array}{ccc}
 a_0^0-\sigma  \sin \left(\theta_s\right)+f_0(980) \cos \left(\theta_s\right) & \sqrt{2} a_0^+ & \sqrt{2} \kappa ^+ \\
 \sqrt{2} a_0^- & -a_0^0-\sigma  \sin \left(\theta_s\right)+f_0(980) \cos \left(\theta_s\right) & \sqrt{2} \kappa  \\
 \sqrt{2} \kappa ^- & \sqrt{2} \bar{\kappa } & \sqrt{2} \sigma  \cos \left(\theta_s\right)+\sqrt{2} f_0(980) \sin \left(\theta_s\right) \\
\end{array}
\right),
\end{equation}
with $\theta_{s} = -21^{\circ}$ being the scalar mixing angle, and $\kappa$ is also known as $K_{0}^{*}(700)$. The interaction coefficients have been fixed to describe the $\pi \pi$ and $\pi K$ scattering data, and also to reproduce the decay $\eta' \to \eta\pi\pi$. 

These studies defined $\eta,\eta'$ in terms of the isoscalar and singlet particles $\eta_{8},\eta_{0}$ in a different way compared to the ALP studies. Namely, the mixing angle in~\cite{Fariborz:1999gr} differs from the angle $\theta_{\eta\eta'} = \arcsin(-1/3)$, commonly used when describing the mixing of ALPs with these mesons. Thus, incorporating the interactions from~\cite{Black:1998wt,Fariborz:1999gr} into the ALP phenomenology ``out of the box'' would lead to inconsistency.

To fix the problem, we start with Eq.~(B.4) of~\cite{Black:1998wt}. It is written in terms of phenomenological constants $A,B,C,D$ and the matrix of the pseudoscalar nonet $\Sigma$ (denoted as $\xi$ there). As usual, we replace it with Eq.~\eqref{eq:sigma-transformed-supplemental}, and perform the expansion in terms of $\mathcal{P}$. The parameters $A,B,C,D$ are fitted to the observational data -- $\pi,K$ scattering for $A,B$, and $\eta'$ decays for $C,D$. Due to the different definition of $\eta,\eta'$ we use, in our expansion, the $\eta,\eta'$ couplings depend differently on $C,D$ than in the expansion (A.1) of Ref.~\cite{Fariborz:1999gr}, so we cannot just use their reported values -- it would destroy the predictions on $\eta'\to \eta\pi\pi$ decay. Instead, we recompute the couplings of scalar mesons to $\eta,\eta'$ in our expansion, and set their values to recover the width of the process $\eta'\to \eta2\pi$, consistent with the experimental value~\cite{ParticleDataGroup:2024cfk}.

We also utilize these couplings when adding the $S$-wave amplitude to the ALP decays into $KK\pi$ using the BaBar fit from~\cite{Cheng:2021kjg}.

\subsubsection{Vector mesons}
\label{app:vector}

The Hidden Local Symmetry approach of including the interaction of vector mesons adds the following Lagrangian~\cite{Fujiwara:1984mp,Guo:2011ir}:
    \begin{align}
    \mathcal{L}_{\text{vec+an}} = & -\frac{3g^{2}}{8\pi^{2}f_{\pi}}\epsilon^{\mu\nu\alpha\beta}\text{Tr}[\mathcal{P}(x)\partial_{\mu}V_{\nu}(x)\partial_{\alpha}V_{\beta}(x)]+\frac{7}{60\pi^{2}f_{\pi}^{5}}\epsilon^{\mu\nu\alpha\beta}\text{Tr}[\mathcal{P}(x)\partial_{\mu}\mathcal{P}\partial_{\nu}\mathcal{P}\partial_{\alpha}\mathcal{P}\partial_{
    \beta}\mathcal{P}]\\ & +2f_{\pi}^{2}\text{Tr}\left|gV_{\mu}-eA_{\mu}Q-\frac{i}{2f_{\pi}^{2}}[\mathcal{P},\partial_{\mu}\mathcal{P}]\right|^{2}
    \label{eq:lagr-vector}
\end{align}
Here, $g \approx m_{\rho}/\sqrt{2}f_{\pi}$, $Q = \text{diag}(2/3,-1/3,-1/3)$ is the quark charge matrix, $V_{\mu}$ is the matrix of vector mesons, 
\begin{equation}
    V_{\mu} = \frac{1}{\sqrt{2}}\begin{pmatrix}
        \frac{\rho^{0}+\omega}{\sqrt{2}} & \rho^{+} & K^{*+} \\ \rho^{-} & \frac{-\rho^{0}+\omega}{\sqrt{2}} & K^{*0} \\ K^{*-} & \bar{K}^{*0} & \phi
    \end{pmatrix},
    \label{eq:vector-matrix}
\end{equation}
with $K^{*}$ being associated with $K^{*}(892)$, and $A_{\mu}$ is the EM field. As discussed, to define the ALP interactions, we start with the corresponding Lagrangian, replace the $\mathcal{P}$ matrix with Eq.~\eqref{eq:P-transformed}, and perform the diagonalization~\eqref{eq:diagonalization}.

The Lagrangian induces the mixing between $A_{\mu}$ and $\rho^{0},\omega,\phi$, which effectively generates, e.g., electromagnetic decays of $P^{0}$. This way, it is responsible for all the vertices of interactions, in particular for the $\kappa_{q}$ cancellation. In particular, it includes the effect of the chiral rotation~\eqref{eq:rotation} coming from the non-invariance of the fermion path integration measure, i.e., reproduces the chiral anomaly in the language of the quark bound states rather than quarks themselves.

Let us show how the $\kappa_{q}$ dependence vanishes by calculating the coupling of the ALPs to two photons. Expanding Eq.~\eqref{eq:lagr-vector}, for the $a\gamma\gamma$ piece, we get
\begin{equation}
    \mathcal{L}_{a\gamma\gamma} = \frac{\alpha_{\text{EM}}\epsilon}{\pi f_{\pi}}c_{\gamma\gamma}^{\text{eff}}F_{\mu\nu}\tilde{F}^{\mu\nu},
    \label{eq:photon-Lagrangian}
\end{equation}
where 
\begin{equation}
    c_{\gamma\gamma}^{\text{eff}} = \frac{1}{9}(6c_{G}(3\kappa_{u}+1)-4\sqrt{6}\theta_{\eta a}-7\sqrt{3}\theta_{\eta'a}-9\theta_{\pi^{0}a})
    \label{eq:cAA-eff}
\end{equation}
The $\kappa_{q}$ dependence vanishes after inserting the expressions~\eqref{eq:mixing-pi0-supplemental}-\eqref{eq:mixing-etapr-supplemental}. The first term has the same structure as the term $\propto \text{Tr}[\kappa_{q} Q^{2}]$, which is explicitly added in the literature~\cite{Bauer:2021wjo} because of the path integral measure non-invariance under the chiral rotation.

We perform two cross-checks. First, in the limit of the ``$P^{0}$-like'' ALP (i.e., setting $\theta_{P^{0'}a} = \delta_{P^{0'}P^{0}}$ and $c_{G} = 0$), the Lagrangian~\eqref{eq:photon-Lagrangian} predicts the decay widths of $\pi^{0},\eta,\eta'$ mesons matching the observed data with $\simeq 10\%$ accuracy, typical for the model of Hidden Local Symmetry~\cite{Fujiwara:1984mp,Ilten:2018crw}.

Second, let us consider the 2-flavor limit. This is achieved by setting $\theta_{\eta/\eta' a}\to 0$, dropping the $\pi^{0}$-$\eta/\eta'$ mixing terms in Eq.~\eqref{eq:mixing-eta-supplemental}, and assuming $\kappa_{u} = 1-\kappa_{d}$ (this is the relation between $\kappa_{u},\kappa_{d}$ in the case of the 2-flavor setup). Keeping now explicitly the ALP derivative coupling to quarks $\mathcal{L}\ni \partial_{\mu}a/f_{a}\ \sum_{q}c_{q}\bar{q}\gamma^{\mu}\gamma_{5}q$, for the expression~\eqref{eq:cAA-eff} we obtain
\begin{equation}
    c_{\gamma\gamma}^{\text{eff}} = c_G \left(-\delta\frac{m_{\pi^{0}}^2}{m_{\pi^{0}}^2-m_a^2}-\frac{5}{3}\right)-\frac{m_a^2 \left(c_u-c_d\right)}{m_{\pi^{0}}^2-m_a^2},
\end{equation}
which exactly matches Eq.~(92) in Ref.~\cite{Bauer:2020jbp} (after taking into account the relation $c_{qq}\equiv 2c_{q}$ there).

\subsubsection{$s\to d$ transition operator}
\label{app:adding-s-to-d}
Next, let us proceed to the process $K\to a \pi$, generated by the $s\to d$ transition. The Standard Model analog of this process, $K\to \pi \pi$, receives contributions from two operators~\cite{Cirigliano:2011ny} classified by the transformation properties of the chiral operators -- $SU(3)$ octet and $27$-plet. The coupling constant in front of the latter is severely suppressed, $G_{27}/G_{8} \approx 0.05$. Despite this, the SM decay is driven by $G_{27}$, as the $G_{8}$ contribution is proportional to the tiny factor $(m_{\pi^{+}}^{2}-m_{\pi^{0}}^{2})/m_{K}^{2}$. However, this is no longer the case for ALPs, and the octet typically makes the main contribution (see, e.g.,~\cite{Gori:2020xvq}).

We implement the octet operator; the implementation of the 27-plet may be, in principle, done analogously. The matrix element of the process has the form
\begin{multline}
    \mathcal{M}_{K^{+}\to \pi^{+}a} = \frac{1}{3} i f_{\pi } G_8 \big[6 \epsilon  c_G (-m_a^2 (2 \kappa_d+\kappa_u)+m_{K^+}^2 (\kappa_d+\kappa_u)+m_{\pi^+}^2 \kappa_d + m_a^2-m_{\pi^+}^2)  \\ +\sqrt{3}\theta_{a \eta '} (m_a^2+2 m_{K^+}^2-3 m_{\pi^+}^2)
   +2 \sqrt{6}  \theta_{\eta a} (m_{K^+}^2-m_a^2)+3 \theta_{\pi^{0}a} (m_a^2-m_{\pi^+}^2)\big]   
\end{multline}
Note that, unlike the work~\cite{Bauer:2021wjo}, we do not decouple the $\eta'$ meson, which leads to the qualitative difference in the scaling of the matrix element with the ALP mass. Namely, after inserting the explicit form of the mixing angles (which identically cancels the $\kappa_{q}$ dependence) and working in the limit $\delta = 0$, we get
\begin{equation}
    \mathcal{M}_{K^{+}\to \pi^{+}a} \approx 8 i \epsilon  c_G f_{\pi } G_8 \frac{ \left(m_{K^+}^2-m_{\pi ^+}^2\right)^2}{m_a^2-m_{\eta'}^{2}}
\end{equation}

\subsubsection{Tensor mesons}

We add the following Lagrangian of the interaction of $P$s with the tensor meson $f_{2}$~\cite{Guo:2011ir}:
\begin{equation}
    \mathcal{L}_{T} = - g_{T} \frac{f_\pi^2}{4}\text{Tr}\left[\left(\partial^\mu \Sigma^{\dagger} \partial^\nu \Sigma-\frac{1}{2} g^{\mu \nu} \partial^\alpha \Sigma^{\dagger} \partial_\alpha \Sigma\right) \boldsymbol{f_{a}}_{\mathbf{2}}\right] f_{2\mu \nu}, 
\end{equation}
where $\boldsymbol{f_{a}}_{\mathbf{2}} = \text{diag}(1/2,1/2,0)$ is the SU(3) generator of the tensor meson. The coupling $g_{T}=13.1 \text{ GeV}^{-1}$ recovers the decay widths of the $f_{2}$ meson~\cite{Cheng:2021kjg}. 

Here, we use the replacement~\eqref{eq:sigma-transformed-supplemental} to achieve the $\kappa_{q}$ independence. As an illustration, consider the vertex $f_{2}a\eta$:
\begin{equation}
    V^{\mu\nu}_{f_{2}a\eta} = \frac{1}{6} \epsilon  g_T\left(-g^{\mu \nu } \left(p_{a}\cdot p_{\eta}\right)+p_{a}^{\mu}
   p_{\eta}^{\nu}+p_{a}^{\nu} p_{\eta}^{\mu}\right)\left(\sqrt{2} \theta _{a \eta '}+\sqrt{6} c_G \left(\kappa _d+\kappa _u\right)+2 \theta _{\text{$\eta $a}}\right)
\end{equation}
Inserting the explicit expressions~\eqref{eq:mixing-eta-supplemental},~\eqref{eq:mixing-etapr-supplemental}, for the $c_{G}$-dependent piece of the coefficient in front of the tensor structure, we get
\begin{equation}
    c_{G}\frac{\left(m_{\eta}^2-m_{\pi^{0}}^{2}\right) \left(m_{a}^2-2 m_{\eta}^{2}+m_{\pi^{0}}^2\right)}{\left(m_{a}^2-m_{\eta}^2\right) \left(m_{a}^2-4 m_{\eta}^2+3 m_{\pi^{0}}^2\right)}
\end{equation}

\subsection{Interaction with nucleons}
\label{app:nucleons}
We follow Ref.~\cite{Blinov:2021say} and consider the low-momentum transfer ALP-nucleon interaction in the form
\begin{equation}
    \mathcal{L}_{app} = g_{app}\partial_{\mu}a\bar{p}\gamma^{\mu}\gamma_{5}p
\end{equation}
Here,
\begin{equation}
    g_{app} = \frac{\left(4 D_s+3 D-F\right) \theta_{a \eta '}}{2 \sqrt{3}}+\epsilon  c_G (\kappa_d (F-D)+2 F \kappa_u)+\frac{\theta_{\eta a} \left(D_s+2 F\right)}{\sqrt{6}}+\frac{1}{2}
   (D+F) \theta_{\pi^{0} a}
\end{equation}
is the ALP-nucleon coupling, with $D_{s},F,D$ being phenomenological constants. Using Eqs.~\eqref{eq:mixing-pi0-supplemental}-\eqref{eq:mixing-etapr-supplemental} and also the relation~\eqref{eq:masses-relations}, the explicit form of $g_{app}$ becomes
\begin{multline}
    g_{app} = \frac{\epsilon  c_G}{f_{\pi}}\frac{ \left(m_{\pi^{0}}^2-m_{\eta }^2\right) \left(m_a^2 \left(3 D_s+2 D\right)-2 m_{\eta }^2 \left(2 D_s+D+F\right)+m_{\pi^{0}}^2 \left(D_s+2 F\right)\right)}{\left(m_a^2-m_{\eta }^2\right) \left(m_a^2-m_{\eta
   '}^2\right)} \\ +\delta \frac{\epsilon  c_G}{f_{\pi}} \frac{ m_{\pi^{0}}^2 (D+F) \left(m_{\pi^{0}}^2-m_{\eta }^2\right) \left(m_a^2-2 m_{\eta }^2+m_{\pi^{0}}^2\right)}{\left(m_{\pi^{0}}^2-m_a^2\right) \left(m_a^2-m_{\eta }^2\right) \left(m_a^2-m_{\eta '}^2\right)}
\end{multline}

\subsection{Adding axial-vector and heavy pseudoscalar excitations}
\label{app:heavy-P-AV}

In the mass range $m_{a}\gtrsim 1\text{ GeV}$, axial-vector mesons $A_{\mu}$ and heavy pseudoscalar excitations $P_{h}$ are essential to describe the ALP interactions. The reason is that ALPs mix with them. For excited pseudoscalar mesons, the mixing structure is similar to the light pseudoscalar case discussed in Sec.~\ref{app:ALP-minimal-ChPT}: the chiral rotation modifies the meson kinetic and mass terms, whereas the derivative coupling translates to the analog of the last term of Eq.~\eqref{eq:lagr-chpt-supplemental}. The axial-vector sector generically adds the mixing terms of the type $L \ni M_{AP}A_{\mu}\partial^{\mu}P$, where $M_{AP}$ is the mixing coefficient.  

It turns out that the $A-a$ mixing is not as important as the mixing with pseudoscalar mesons: unlike the latter, it does not get resonantly enhanced in the vicinity of the $A$ mass. To understand this feature, let us treat the $A_{\mu}\partial^{\mu}a$ term as the interaction term and use it as an internal vertex of some matrix element $\mathcal{M}$ of the ALP production or decay. It has the form
\begin{equation}
    \mathcal{M} \propto M_{A^{0}a}p^{\mu}_{a}D_{\mu\nu}^{A^{0}} \tilde{\mathcal{M}}^{\nu} = M_{A^{0}a}\frac{\left(1-\frac{m_{a}^{2}}{m_{A^{0}}^{2}}\right)}{m_{a}^{2}-m_{A^{0}}^{2}-i\Gamma_{A^{0}}m_{A^{0}}} p^{\nu}\tilde{\mathcal{M}}_{\nu} \equiv \theta_{aA^{0}}^{\text{eff}}p^{\nu}\tilde{\mathcal{M}}_{\nu},
    \label{eq:mixing-axial-vector}
\end{equation}
where $\theta_{aA^{0}}^{\text{eff}}$ is the mixing angle. We see that it vanishes at $m_{a} = m_{A^{0}}$, and hence lacks the enhancement. Therefore, compared to the mixing with the heavy pseudoscalar mesons, it is expected to be subdominant. In addition, incorporating them in a consistent way with the other mesons that are already included in our approach is a non-trivial task. For these two reasons, we will drop their interactions with ALPs. We return to this question in Sec.~\ref{app:open-problems}.

Nevertheless, as we will see, the mixing between axial-vector and light pseudoscalar mesons is important to describe the interactions of pseudoscalar excitations with vector and light pseudoscalar mesons. Therefore, we briefly describe their phenomenology in Sec.~\ref{app:adding-VA}.

Our treatment of their interactions follows Refs.~\cite{Parganlija:2012fy,Parganlija:2016yxq,Giacosa:2024epf}. It utilizes the description of different vector, axial-vector, heavy scalar, and pseudoscalar excitations using the $SU(3)$ symmetry, assuming that masses originate from a condensate of the heavy scalar nonet, and fixing the interaction couplings by matching the theoretical prediction of various decay widths with experimentally observed data. This model is called the \textit{E}xtended \textit{L}inear \textit{S}igma Model, or \textbf{ELSM}. The crucial point is that we turn off the contributions from axial-vector and heavy pseudoscalar sectors for the scale $m \lesssim 1\text{ GeV}$, which is explained by the incompleteness of the ELSM in the meson spectrum mass range $M < 1\text{ GeV}$ (see Sec.~\ref{app:open-problems}).

The mixing of ALPs with $P_{h}$ contributes to all ``$P_{h}$-like'' decay modes, including $3\pi, KK\pi, \eta2\pi$, and many others.

\subsubsection{Mixing of axial-vector and pseudoscalar mesons}
\label{app:adding-VA}

In this section, we follow the approach of including the axial-vector mesons presented in Ref.~\cite{Parganlija:2012fy}.

We start with the Lagrangian describing the sector of axial-vector $A$, vector $V$, and pseudoscalar mesons $P$:
\begin{align}
\mathcal{L}_{0}  &
=\mathop{\mathrm{Tr}}[(D_{\mu}\Phi)^{\dagger}(D_{\mu}\Phi)]-m_{0}^{2}\left(
\frac{G}{G_{0}}\right)  ^{2}\mathop{\mathrm{Tr}}(\Phi^{\dagger}\Phi
)-\lambda_{1}[\mathop{\mathrm{Tr}}(\Phi^{\dagger}\Phi)]^{2}-\lambda
_{2}\mathop{\mathrm{Tr}}(\Phi^{\dagger}\Phi)^{2}{\nonumber}\nonumber\\
&  -\frac{1}{4}\mathop{\mathrm{Tr}}(L_{\mu\nu}^{2}+R_{\mu\nu}^{2}%
)+\mathop{\mathrm{Tr}}\left[  \left(  \left(  \frac{G}{G_{0}}\right)
^{2}\frac{m_{1}^{2}}{2}+\Delta\right)  (L_{\mu}^{2}+R_{\mu}^{2})\right]
+\mathop{\mathrm{Tr}}[H(\Phi+\Phi^{\dagger})]+\mathop{\mathrm{Tr}}(\Phi
^{\dagger}\Phi E_{0}+\Phi\Phi^{\dagger}E_{0}){\nonumber}\nonumber\\
&  +c_{1}(\det\Phi-\det\Phi^{\dagger})^{2}+i\frac{g_{2}}{2}%
(\mathop{\mathrm{Tr}}\{L_{\mu\nu}[L^{\mu},L^{\nu}%
]\}+\mathop{\mathrm{Tr}}\{R_{\mu\nu}[R^{\mu},R^{\nu}]\}){\nonumber}\nonumber\\
&  +\frac{h_{1}}{2}\mathop{\mathrm{Tr}}(\Phi^{\dagger}\Phi
)\mathop{\mathrm{Tr}}(L_{\mu}^{2}+R_{\mu}^{2})+h_{2}%
\mathop{\mathrm{Tr}}[\vert L_{\mu}\Phi \vert ^{2}+\vert \Phi R_{\mu} \vert ^{2}]+2h_{3}%
\mathop{\mathrm{Tr}}(L_{\mu}\Phi R^{\mu}\Phi^{\dagger}){\nonumber}\nonumber\\
&  +g_{3}[\mathop{\mathrm{Tr}}(L_{\mu}L_{\nu}L^{\mu}L^{\nu}%
)+\mathop{\mathrm{Tr}}(R_{\mu}R_{\nu}R^{\mu}R^{\nu})]+g_{4}%
[\mathop{\mathrm{Tr}}\left(  L_{\mu}L^{\mu}L_{\nu}L^{\nu}\right)
+\mathop{\mathrm{Tr}}\left(  R_{\mu}R^{\mu}R_{\nu}R^{\nu}\right)
]{\nonumber}\nonumber\\
&  +g_{5}\mathop{\mathrm{Tr}}\left(  L_{\mu}L^{\mu}\right)
\,\mathop{\mathrm{Tr}}\left(  R_{\nu}R^{\nu}\right)  +g_{6}%
[\mathop{\mathrm{Tr}}(L_{\mu}L^{\mu})\,\mathop{\mathrm{Tr}}(L_{\nu}L^{\nu
})+\mathop{\mathrm{Tr}}(R_{\mu}R^{\mu})\,\mathop{\mathrm{Tr}}(R_{\nu}R^{\nu
})]\text{ .}     \label{eq:VA}%
\end{align}
Here, 
\begin{equation}
    \Phi(x) = \langle \Phi\rangle +S_{h}+ i\mathcal{P}(x)
    \label{eq:phi-nonet}
\end{equation} 
is the (heavy scalar)-pseudoscalar nonet, with $S_{h}$ being the heavy scalar nonet, and 
\begin{equation}
    \langle \Phi\rangle = \frac{1}{\sqrt{2}}\text{diag}\left(\frac{\phi_{N}}{\sqrt{2}}, \frac{\phi_{N}}{\sqrt{2}}, \phi_{S} \right)
    \label{eq:phi-matrix}
\end{equation}
being the scalar condensate giving mass to the fields. 
\begin{equation}
    L_{\mu} = V_{\mu}+A_{\mu}, \quad R_{\mu} = V_{\mu}-A_{\mu},
\end{equation}
are, correspondingly, left and right axial and vector nonets, with $V_{\mu}$ defined in Eq.~\eqref{eq:vector-matrix}, and $A_{\mu}$ the axial-vector nonet:
\begin{equation}
A_{\mu} = \frac{1}{\sqrt{2}}\begin{pmatrix} \frac{f_{1N}+a_{1}^{0}}{\sqrt{2}} & a_{1}^{+} & K_{1}^{+} \\ 
 a_{1}^{-}& \frac{f_{1N}-a_{1}^{0}}{\sqrt{2}} & K_{1}^{0}\\ K^{-}_{1}  & \bar{K}^{0}_{1} & f_{1S}  \end{pmatrix}
\end{equation}
$f_{1N}$ is associated to $f_{1}(1285)$, $f_{1S}$ to $f_{1}(1415)$, and $K_{1}$ to $K_{1}(1270)$. $L_{\mu\nu} = \partial_{\mu}L_{\nu} - \partial_{\nu}R_{\mu}$ (and similarly $R_{\mu\nu}$) is the field strength. Finally, $D_{\mu}\Phi = \partial_{\mu}\Phi - ig_{1}(L_{\mu}\Phi - \Phi R_{\mu})$ is the covariant derivative, with $g_{1}$ being an interaction coupling.

The relevant values of the couplings entering Eq.~\eqref{eq:VA} are summarized in Table~III of Ref.~\cite{Parganlija:2012fy}; the rest are set to zero. Finally, the glueball $G$ is set to its VEV value $G_{0}$, whereas $E_{0} = \text{diag}(0,0,\epsilon_{S})$. 

We are interested in the mixing terms between $A$ and $P$ particles. Let us therefore turn off vector and heavy scalar fields $S_{h}$.\footnote{We have a posteriori checked that their large mass, absence of mixing with ALPs, and presence of the lighter scalar nonet make their contribution to the processes with ALPs subdominant.} Thus, the matrix $\Phi(x)$ has the form $\Phi(x) = \langle \Phi\rangle + iP(x)$. The quadratic part of the resulting interaction is
\begin{equation}
    \mathcal{L}_{VA}^{(2)} = \sum_{A,X = \{P,a\}} \frac{M_{AX}}{2}A_{\mu}\partial^{\mu}X
    \label{eq:VA-quadratic}
\end{equation}
With this Lagrangian, we reproduce the Lagrangian of the pure $A$-$P$ mixing from Ref.~\cite{Parganlija:2012fy}. 

It is possible to eliminate the mixing terms by performing the shift
\begin{equation}
    A^{\mu} \to A^{\mu}+M_{A^{0}P}Z_{P}\partial^{\mu}P, \quad P\to Z_{P}P
    \label{eq:VA-shift}
\end{equation}
where $Z_{P}$ is the renormalization constant, which ensures that the kinetic term for $P$ has the canonical form after the shift. In the ELSM, $Z_{P}$ enters the expressions of the $P$ masses in terms of the parameters of the model~\eqref{eq:VA}, as well as the $P$s' interactions with other mesons. 

Our approach to treating this mixing is to assume that the Lagrangian~\eqref{eq:VA} and the shift~\eqref{eq:VA-shift} generate the ChPT mass term~\eqref{eq:pure-ChPT}. I.e., in the sector of pseudoscalar mesons only, the ELSM predictions exactly match the ChPT description. However, we will utilize it in the next section, where we describe the interactions of the heavy pseudoscalar nonet.

For completeness, however, we provide the mixing coefficients between the neutral $A^{0}$ and ALPs, which can be obtained if including them in the same fashion as in the other interactions:
\begin{align}
    M_{a^{0}_{1}a} =& \ g_1 \phi_N \left(c_G \left(\kappa_d-\kappa_u\right)-\theta_{\pi^{0}a}\right), \\ M_{f_{1}(1285)a} =& \ -\frac{1}{3} g_1 \phi_N \left(\sqrt{3} \theta_{\eta'a }+3 c_G \left(\kappa_d+\kappa_u\right)+\sqrt{6} \theta_{\eta a}\right), \\ M_{f_{1}(1415)a} =& \ \frac{1}{3} g_1 \phi_S \left(-2 \sqrt{3}\theta_{\eta'a}+6 c_G\left(\kappa_d+\kappa_u-1\right)+\sqrt{6} \theta_{\eta a}\right) 
\end{align}
As it should be, it is $\kappa_{q}$-independent; it also does not vanish in the limit $c_{u,d,s}\to 0$.

\subsubsection{Heavy pseudoscalar sector}
\label{app:adding-Ph}
To describe the interactions of ALPs with heavy pseudoscalar mesons, we follow Ref.~\cite{Parganlija:2016yxq}, which incorporated them in the ELSM. The starting Lagrangian is
\begin{align}
\mathcal{L}_{P_{h}}  &
=\mathop{\mathrm{Tr}}[(D_{\mu}\Phi_{h})^{\dagger}(D_{\mu}\Phi_{h})]+\alpha
\mathop{\mathrm{Tr}}[(D_{\mu}\Phi_{h})^{\dagger}(D_{\mu}\Phi) + (D_{\mu}\Phi)^{\dagger}(D_{\mu}\Phi_{h})]-(m_{0}%
^{\ast})^{2}\left(  \frac{G}{G_{0}}\right)  ^{2}\mathop{\mathrm{Tr}}(\Phi
_{h}^{\dagger}\Phi_{h})\nonumber\\
&  -\lambda_{0}\left(  \frac{G}{G_{0}}\right)  ^{2}\mathop{\mathrm{Tr}}(\Phi
_{h}^{\dagger}\Phi+\Phi^{\dagger}\Phi_{h})-\lambda_{1}^{\ast}%
\mathop{\mathrm{Tr}}(\Phi_{h}^{\dagger}\Phi_{h})\mathop{\mathrm{Tr}}(\Phi
^{\dagger}\Phi)-\lambda_{2}^{\ast}\mathop{\mathrm{Tr}}(\Phi_{h}^{\dagger}%
\Phi_{h}\Phi^{\dagger}\Phi+\Phi_{h}\Phi_{h}^{\dagger}\Phi\Phi^{\dagger
})\nonumber\\
&  -\kappa_{1}\mathop{\mathrm{Tr}}(\Phi_{h}^{\dagger}\Phi+\Phi^{\dagger}%
\Phi_{h})\mathop{\mathrm{Tr}}(\Phi^{\dagger}\Phi)-\kappa_{2}%
[\mathop{\mathrm{Tr}}(\Phi_{h}^{\dagger}\Phi+\Phi^{\dagger}\Phi_{h}%
)]^{2}-\kappa_{3}\mathop{\mathrm{Tr}}(\Phi_{h}^{\dagger}\Phi+\Phi^{\dagger
}\Phi_{h})\mathop{\mathrm{Tr}}(\Phi_{h}^{\dagger}\Phi_{h})-\kappa
_{4}[\mathop{\mathrm{Tr}}(\Phi_{h}^{\dagger}\Phi_{h})]^{2}\nonumber\\
&  -\xi_{1}\mathop{\mathrm{Tr}}(\Phi_{h}^{\dagger}\Phi\Phi^{\dagger}\Phi
+\Phi_{h}\Phi^{\dagger}\Phi\Phi^{\dagger})-\xi_{2}\mathop{\mathrm{Tr}}(\Phi
_{h}^{\dagger}\Phi\Phi_{h}^{\dagger}\Phi+\Phi^{\dagger}\Phi_{h}\Phi^{\dagger
}\Phi_{h})-\xi_{3}\mathop{\mathrm{Tr}}(\Phi^{\dagger}\Phi_{h}\Phi_{h}%
^{\dagger}\Phi_{h}+\Phi\Phi_{h}^{\dagger}\Phi_{h}\Phi_{h}^{\dagger})-\xi
_{4}\mathop{\mathrm{Tr}}(\Phi_{h}^{\dagger}\Phi_{h})^{2}{\nonumber}\nonumber\\
&  +\mathop{\mathrm{Tr}}(\Phi_{h}^{\dagger}\Phi_{h}E_{1}+\Phi_{h}\Phi
_{h}^{\dagger}E_{1})+c_{1}^{\ast}[(\det\Phi-\det\Phi_{h}^{\dagger})^{2}%
+(\det\Phi^{\dagger}-\det\Phi_{h})^{2}]+c_{1E}^{\ast}(\det\Phi_{h}-\det
\Phi_{h}^{\dagger})^{2}{\nonumber}\nonumber\\
&  +\frac{h_{1}^{\ast}}{2}\mathop{\mathrm{Tr}}(\Phi_{h}^{\dagger}\Phi
+\Phi^{\dagger}\Phi_{h})\mathop{\mathrm{Tr}}(L_{\mu}^{2}+R_{\mu}^{2}%
)+\frac{h_{1E}^{\ast}}{2}\mathop{\mathrm{Tr}}(\Phi_{h}^{\dagger}\Phi
_{h})\mathop{\mathrm{Tr}}(L_{\mu}^{2}+R_{\mu}^{2}){\nonumber}\nonumber\\
&  +h_{2}^{\ast}\mathop{\mathrm{Tr}}(\Phi_{h}^{\dagger}L_{\mu}L^{\mu}\Phi
+\Phi^{\dagger}L_{\mu}L^{\mu}\Phi_{h}+R_{\mu}\Phi_{h}^{\dagger}\Phi R^{\mu
}+R_{\mu}\Phi^{\dagger}\Phi_{h}R^{\mu})+h_{2E}^{\ast}%
\mathop{\mathrm{Tr}} [\vert L_{\mu}\Phi_{h} \vert ^{2}+\vert \Phi_{h} R_{\mu} \vert ^{2}]{\nonumber}\nonumber\\
&  +2h_{3}^{\ast}\mathop{\mathrm{Tr}}(L_{\mu}\Phi_{h}R^{\mu}\Phi^{\dagger
}+L_{\mu}\Phi R^{\mu}\Phi_{h}^{\dagger})+2h_{3E}^{\ast}%
\mathop{\mathrm{Tr}}(L_{\mu}\Phi_{h}R^{\mu}\Phi_{h}^{\dagger})\text{ .}
\label{eq:heavy-pseudovector-supplemental} %
\end{align}
Here,
\begin{equation}
    \Phi_{h} = S_{hh} + \frac{i}{\sqrt{2}}\begin{pmatrix} \frac{\eta_{N}+\pi^{0}_{E}}{\sqrt{2}}& \pi^{+}_{E} & K^{+}_{E} \\ \pi^{-}_{E} & \frac{\eta_{N}-\pi^{0}_{E}}{\sqrt{2}}  & K^{0}_{E} \\ K^{-}_{E} & \bar{K}^{0}_{E} & \eta_{N} \end{pmatrix},
\end{equation}
is the excited scalar-pseudoscalar nonet, with $S_{hh}$ denoting the second heavy scalar fields. Here, $\pi_{E}$ is associated to $\pi(1300)$, $K_{E}$ to $K(1460)$, $\eta_{N}$ to $\eta(1295)$ and $\eta_{S}$ to $\eta(1440)$. The association assumes that the mentioned states are 2-quark bound states, although there is ambiguity. As in the axial-vector sector, the covariant derivative has the form $D_{\mu}\Phi_{h} = \partial_{\mu}\Phi_{h} - g_{1E}(L_{\mu}\Phi_{h} - \Phi_{h}R_{\mu})$, and $G = G_{0}$.

Following Ref.~\cite{Parganlija:2016yxq}, we drop the terms proportional to $\alpha, \lambda_{0}^{*},\lambda_{1}^{*},\xi_{1}, h_{1}^{*},\kappa_{1-4}$. It implies, in particular, the absence of mixing between the heavy and light pseudoscalar mesons. We comment on its consequences in Sec.~\ref{app:open-problems}. We also set $S_{hh} = 0$, as similarly to the nonet $S_{h}$, they are not expected to significantly affect the interactions.

The interactions of $\Phi_{h}$ with vector and light pseudoscalar mesons are obtained with the help of the transformation~\eqref{eq:VA-shift}; it induces the vertices $\Phi_{h}\Phi V$, $\Phi_{h}\Phi \Phi \Phi$, dominating the decays of $\Phi_{h}$ and hence ALPs, due to the mixing with heavy pseudoscalars.

After inserting the explicit form of $\Phi$ matrix, Eq.~\eqref{eq:phi-nonet}, one arrives at the following mass and kinetic matrices between the ALP and the neutral mesons $P_{h}^{0} = \pi^{0}(1300),\eta(1295), \eta(1440)$, similar to Eqs.~\eqref{eq:kinetic-matrix},~\eqref{eq:mass-matrix}:
{\small
\begin{equation}
\hat{M} = \left(
\begin{array}{cccc}
 m_a^2 & M_{a\pi^{0}(1300)} & M_{a\eta(1295)} & M_{a\eta(1440)} \\ M_{a\pi^{0}(1300)} & m_{\pi^{0}(1300)}^2 & 0 & 0 \\
  M_{a\eta(1295)} & 0 & m_{\eta(1295)}^2 & 0 \\
M_{a\eta(1440)} & 0 & 0 & m_{\eta(1440)}^2 \\
\end{array}
\right)
    \label{eq:mass-matrix-heavy}
\end{equation}
}
with
\begin{align}
    M_{a\pi^{0}(1300)} &= -c_{G}\frac{f_{\pi}}{2f_{a}}\left(\kappa_d-\kappa_u\right) \left(2 m^{*2}_{0}+\left(\lambda^{*}_{2}-\xi_2\right) \phi_N^2\right), \\  M_{a\eta(1295)} &=c_{G}\frac{f_{\pi}}{2f_{a}}\left(\kappa_d+\kappa_u\right) \left(2m^{*2}_{0}+\left(\lambda^{*}_{2}-\xi_2\right) \phi_N^2\right), \\ M_{a\eta(1440)} &= -c_G \frac{f_{\pi}}{f_{a}} \sqrt{2} \left(\kappa_d+\kappa_u-1\right) \left(m^{*2}_{0}+\left(\lambda^{*}_{2}-\xi_2\right) \phi_S^2-2\epsilon^{E}_{S}\right),
\end{align}
and
\begin{equation}
\hat{K}= \left(
\begin{array}{cccc}
 1 & \frac{f_{\pi } \left(c_G \left(\kappa _u-\kappa _d\right)-c_d+c_u\right)}{f_{a}} & \frac{f_{\pi } \left(c_G \left(\kappa _d+\kappa _u\right)+c_d+c_u\right)}{f_{a}} & \frac{\sqrt{2} f_{\pi } \left(c_{s}-c_G \left(\kappa
   _d+\kappa _u-1\right)\right)}{f_{a}} \\
 \frac{f_{\pi } \left(c_G \left(\kappa _u-\kappa _d\right)-c_d+c_u\right)}{f_{a}} & 1 & 0 & 0 \\
 \frac{f_{\pi } \left(c_G \left(\kappa _d+\kappa _u\right)+c_d+c_u\right)}{f_{a}} & 0 & 1 & 0 \\
 \frac{\sqrt{2} f_{\pi } \left(c_{s}-c_G \left(\kappa _d+\kappa _u-1\right)\right)}{f_{a}} & 0 & 0 & 1 \\
\end{array}
\right)
    \label{eq:kinetic-matrix-heavy}
\end{equation}
Unlike the case of the light pseudoscalar sector, there is no mixing between various mesons from the beginning -- the mixing is solely between $P^{0}_{h}$ and $a$. Utilizing the relation between the parameters of the Lagrangian and the masses of the heavy pseudoscalars (with the help of our \textsc{Mathematica} notebook, we recovered Eqs.~(21)-(27) in Ref.~\cite{Parganlija:2016yxq}), for the resulting mixing angles, we get
\begin{align}
    \theta_{\pi^{0}(1300)a} &= -\frac{f_{\pi}}{f_{a}}\left( c_{G}\left(\kappa_u-\kappa_d\right)+\frac{m_a^2 (c_u-c_d)}{m_a^2-m_{\pi^{0}(1300)}^2}\right), \label{eq:mixing-pi01300-supplemental}\\ \theta_{\eta(1295) a} &= -\frac{f_{\pi}}{f_{a}}\left(c_G \left(\kappa_d+\kappa_u\right)+\frac{m_a^2
   \left(c_d+c_u\right)}{m_a^2-m_{\eta(1295)}^2}\right), \label{eq:mixing-eta1295-supplemental}\\ \theta_{\eta(1440)a} &= -\sqrt{2}\frac{f_{\pi}}{f_{a}}\left(c_{G}(1-\kappa_d-\kappa_u)+\frac{m_a^2 c_{s}}{m_a^2-m_{\eta(1440)}^2}\right)
   \label{eq:mixing-eta1440-supplemental}
\end{align}
Unlike the light pseudoscalars, heavy pseudoscalar mesons have a non-negligible decay width, which smears the resonant enhancement. To include this effect in practice, we ensure that the $\kappa_{q}$-dependence vanishes in the zero-width limit, explicitly set $\kappa_{q} = 0$, and replace the denominators of the mixing angles: $m^{2}_{a}-m_{P^{0}_{h}}^{2}\to m^{2}_{a}-m_{P^{0}_{h}}^{2}-i\Gamma_{P^{0}_{h}}m_{P^{0}_{h}}$.

\subsubsection{How the $c_{G}$-mediated contribution vanishes}
Now, let us consider the contribution of the Lagrangian~\eqref{eq:heavy-pseudovector-supplemental} to some processes. Choosing, for instance, the process $a(\to \eta^{*}(1440)) \to KK\pi$, we get for the corresponding Lagrangian
\begin{multline}
    \mathcal{L}_{a} = \frac{if_{\pi}}{2f_{a}}a(x) \partial^{\mu}K^{-}(x) K^{*+}_{\mu}(x) h_3^{*} \theta_{K^{-}_{1}(1270)K^{-}}\bigg[\sqrt{2} \left(\phi_S \left(\theta_{\eta (1295)a}+\theta_{\pi^0(1300)a}\right)-\phi_N \theta_{\eta (1440)a}\right)\\ +2 c_G\left(\phi_N \left(\kappa_d+\kappa_u-1\right)+\sqrt{2} \phi_S \kappa_u\right)\bigg]
\end{multline}
After using Eqs.~\eqref{eq:mixing-pi01300-supplemental}-\eqref{eq:mixing-eta1440-supplemental}, for the expression in the brackets, we get
\begin{equation}
    \frac{2 c_{s} \phi_{N}}{m_{a}^2-m_{\eta(1440)}^{2}}-\frac{\sqrt{2} \phi_{S} (c_{d} (m_{\eta(1295)}^{2}-m_{\pi^{0}(1300)}^{2})+c_{u} (2 m_{a}^2-m_{\eta(1295)}^2-m_{\pi^{0}(1300)}^2))}{(m_{a}^2-m_{\eta(1295)}^2) (m_{a}^2-m^{2}_{\pi^{0}(1300)})}
\end{equation}
I.e., whereas the $c_{u,d,s}$ terms obviously survive, the dependence on $c_{G}$ vanishes identically. It happens because of the simple structure of the interaction operators and the trivial structure of heavy pseudoscalars utilized in Ref.~\cite{Parganlija:2016yxq}: unlike the case of light pseudoscalars, $\eta(1295)$, $\eta(1440)$ exactly match the isoscalar and singlet components of the nonet; also, there is no analog of the isospin breaking.

The same cancellation also applies to any other interaction induced by the Lagrangian~\eqref{eq:heavy-pseudovector-supplemental}. Therefore, if dropping the terms proportional to $\alpha, \lambda_{0}^{*},\lambda_{1}^{*},\xi_{1}, h_{1}^{*},\kappa_{1-4}$ (as it is done in~\cite{Parganlija:2016yxq}), the ALPs coupled to gluons can only experience these interactions via tiny couplings $c_{u,d,s}$ induced because of the renormalization flow of the ALP gluonic operator from the scale $\Lambda$.

\subsection{Cross-checks of our approach}
\label{app:cross-checks}
We have validated our combined approach in various independent ways:
\begin{itemize}
    \item[1.] Comparing symbolic expressions for various quantities with the literature. For instance, assuming the 2-flavor limit, we recovered the mass matrix~\eqref{eq:mass-matrix}, the ALP-photon coupling, and the ALP decay width into three pions with the works~\cite{Bauer:2017ris,Bauer:2020jbp} (see also Sec.~\ref{app:vector}). Next, we have compared the expressions for the pure ChPT contribution to the matrix elements of ALP decays to Ref.~\cite{Aloni:2018vki} and mostly found the exact agreement. The only exception concerns the $\kappa_{q}$-dependent piece, which is absent for the $a\to \eta\pi \pi$ matrix element in~\cite{Aloni:2018vki}. For the added interactions with scalar, vector, and tensor mesons (where we followed Ref.~\cite{Cheng:2021kjg}), we have found the agreement with~\cite{Cheng:2021kjg} except for the sector of the scalar mesons, modified by our treatment (discussed in Sec.~\ref{app:scalar}).
    \item[2.] Ensuring $\kappa_{q}$ independence of various vertices and matrix elements. When computing different quantities, we insert an explicit form of the mixing angles and calculate the $\kappa_{q}$-dependent pieces. They always vanish.
    \item[3.] Comparing the widths of the Standard Model processes with their measured values. We consider the processes $\eta \to \pi^{+}\pi^{-}\gamma$, $\eta' \to \eta2\pi$, $\eta' \to 4\pi$, $\pi^{0}/\eta/\eta'\to 2\gamma$, which in the model we use goes via the mixing of vector mesons with photons~\cite{Fujiwara:1984mp} and intermediate vector and scalar excitations~\cite{Guo:2011ir,Fariborz:1999gr}. We find good agreement within 10\%. This deviation is subdominant compared to the other uncertainty sources (highlighted in Sec.~\ref{app:open-problems}) and also uncertainties coming from the experimental setup (which may include a cascade enhancement of the ALP production in the thick target -- already an $\mathcal{O}(1)$ effect) and lack of knowledge about the meson's spectra. Regarding the newly added axial-vector and heavy pseudoscalar sectors, we have recovered the results of Refs.~\cite{Parganlija:2012fy,Parganlija:2016yxq}.
    \item[4.] Reproducing the results of Ref.~\cite{DallaValleGarcia:2023xhh}, which studied the ALPs universally coupled to fermions, after turning off the contributions from axial-vector and heavy pseudoscalar mesons. For this setup, the hadronic ALP widths differ within a factor of 1.5. The discrepancy is caused by the improvement of the sector of the interactions with the scalar mesons made in the given paper (see Sec.~\ref{app:scalar}).
\end{itemize}

\begin{figure}
    \centering
    \includegraphics[width=0.5\linewidth]{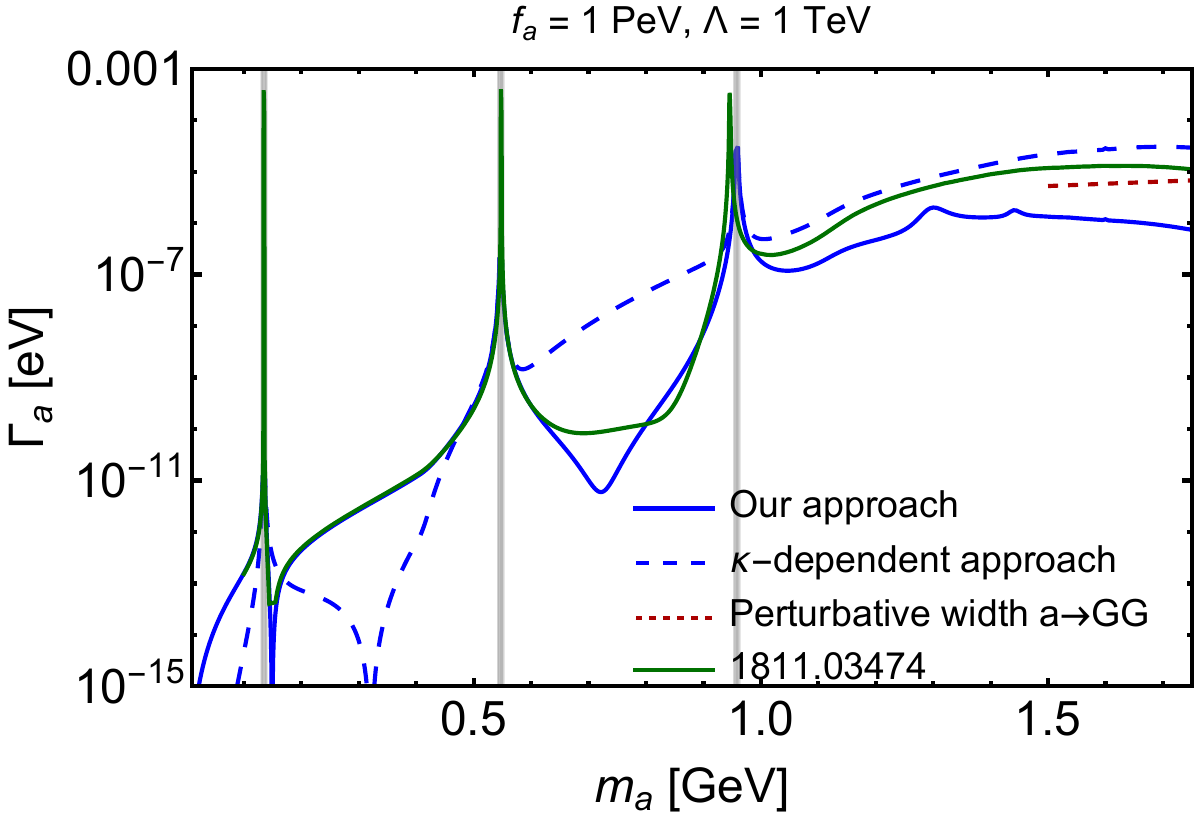}~\includegraphics[width=0.5\linewidth]{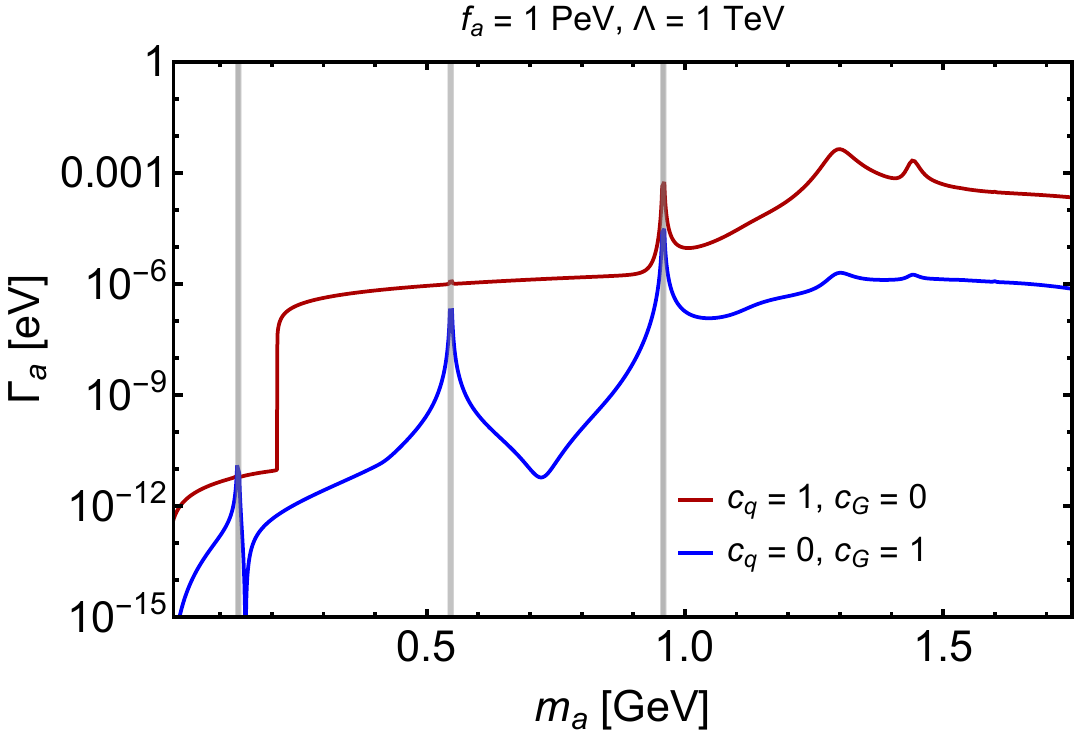}
    \caption{\textit{Left panel}: the comparison of the total ALP decay widths obtained in this work and in Ref.~\cite{Aloni:2018vki}. The meaning of the blue lines and the dashed red line is the same as in Fig.~\ref{fig:production-probabilities} of the main text. \textit{Right panel}: the comparison of the total decay widths of the fermionic and gluonic ALPs. For the fermionic coupling case, the jump at $m_{a}=2m_{\mu}$ is caused by the turning on of the di-muon decay, which dominates until the mass $m_{a} \simeq m_{\eta'}$.}
    \label{fig:widths-comparison}
\end{figure}

Let us finalize the discussion by comparing our calculation of the ALP decay width with the results of Ref.~\cite{Aloni:2018vki}. In Fig.~\ref{fig:widths-comparison}, we show the result of our approach, the $\kappa$-dependent approach without the inclusion of the mixing with heavy excitations, and the total width from Fig. S1 of~\cite{Aloni:2018vki}. For masses $m_{a}\lesssim 0.5\text{ GeV}$, the width from~\cite{Aloni:2018vki} closely matches our results, at masses $m_{\eta}\lesssim m_{a}\lesssim m_{\eta'}$, it deviates from both our curves (still being closer to the $\kappa$-independent approach), whereas at higher masses, it is much closer to the $\kappa$-dependent calculations. This behavior is explained by the fact that at low masses, the ALP width is dominated by the decay into two photons $a\to \gamma\gamma$, for which Ref.~\cite{Aloni:2018vki} used the $\kappa_{q}$-invariant description, manually including the contribution from the chiral rotation (recall a discussion in Sec.~\ref{app:adding-SVT}). For higher masses, hadronic decay modes (for which the $\kappa$-dependent approach has been used in~\cite{Aloni:2018vki}) become more and more important, and the discrepancy quickly accumulates, reaching 1-2 orders of magnitude.

The result of~\cite{Aloni:2018vki} is similar to our calculations obtained within the $\kappa_{q}$-dependent approach (which does not utilize the rotations~\eqref{eq:P-transformed-supplemental},~\eqref{eq:sigma-transformed-supplemental}. The discrepancy arises mainly because of the differences in the treatment of light resonances, which we described in Sec.~\ref{app:ALP-minimal-ChPT}. In the same figure (right panel), we compare the total decay widths of ALPs that are universally coupled to fermions and those coupling solely to gluons. In the fermionic case, the effects of the heavy pseudoscalar resonances are much stronger (see Eq.~\eqref{eq:mixing-eta1440-supplemental}), leading to a significant enhancement of the decay width.

\subsection{Limitations of our approach}
\label{app:open-problems}

Our approach to ALP phenomenology, summarized in Sec.~\ref{app:ALP-interactions}, has limitations intimately related to the maturity of the frameworks describing effective interactions of mesons. There are two major problems.

The first problem arises because there is no unique comprehensive framework describing all the meson excitations with masses below $2\text{ GeV}$~\cite{Giacosa:2024epf}. For instance, ELSM is fitted to describe the main decays of some of the pseudoscalar, scalar, vector, and tensor mesons. On the other hand, it does not consistently incorporate the sectors of light scalar nonet and pseudovector fields, as well as anomalous interactions with vector mesons. The latter may be essential for decays of light pseudoscalar mesons~\cite{Guo:2011ir,Fariborz:1999gr}. Partially because of this, it does not fit the experimental data on various subdominant decays of mesons, which are essential for describing the ALP decay pattern, such as $\eta \to \pi^{+}\pi^{-}\gamma$ or $\eta' \to 4\pi$. Because of this, we use the hybrid framework, utilizing the ELSM to describe the sector of heavy pseudoscalars.

Next, some heavy pseudoscalar excitations, such as $\pi(1800)$, are not implemented as well. Because of the resonantly enhanced mixing of such mesons with the ALPs, it would severely influence the ALP phenomenology. 

One of the main reasons for this is the lack of robust experimental data on these resonances~\cite{ParticleDataGroup:2024cfk}, leading to huge uncertainties in decay widths and masses of some resonances. It leads to a complexity of interpreting various meson states assigned in Sec.~\ref{app:heavy-P-AV} as 2-quark bound states may be 4-quark bound states or an admixture between 2-quark and 4-quark bound states~\cite{Giacosa:2024epf}; the latter would directly change the interaction operators of mesons and hence the ALP phenomenology.

An independent problem concerns the ambiguity of the phenomenological Lagrangians. For example, consider again the interaction Lagrangian of heavy pseudoscalar mesons $P_{h}$ as defined in Ref.~\cite{Parganlija:2016yxq}:
\begin{align}
\mathcal{L}_{P_{h}}  &
=\mathop{\mathrm{Tr}}[(D_{\mu}\Phi_{h})^{\dagger}(D_{\mu}\Phi_{h})]+\alpha
\mathop{\mathrm{Tr}}[(D_{\mu}\Phi_{h})^{\dagger}(D_{\mu}\Phi) + (D_{\mu}\Phi)^{\dagger}(D_{\mu}\Phi_{h})]-(m_{0}%
^{\ast})^{2}\left(  \frac{G}{G_{0}}\right)  ^{2}\mathop{\mathrm{Tr}}(\Phi
_{h}^{\dagger}\Phi_{h})\nonumber\\
&  -\lambda_{0}\left(  \frac{G}{G_{0}}\right)  ^{2}\mathop{\mathrm{Tr}}(\Phi
_{h}^{\dagger}\Phi+\Phi^{\dagger}\Phi_{h})-\lambda_{1}^{\ast}%
\mathop{\mathrm{Tr}}(\Phi_{h}^{\dagger}\Phi_{h})\mathop{\mathrm{Tr}}(\Phi
^{\dagger}\Phi)-\lambda_{2}^{\ast}\mathop{\mathrm{Tr}}(\Phi_{h}^{\dagger}%
\Phi_{h}\Phi^{\dagger}\Phi+\Phi_{h}\Phi_{h}^{\dagger}\Phi\Phi^{\dagger
})\nonumber\\
&  -\kappa_{1}\mathop{\mathrm{Tr}}(\Phi_{h}^{\dagger}\Phi+\Phi^{\dagger}%
\Phi_{h})\mathop{\mathrm{Tr}}(\Phi^{\dagger}\Phi)-\kappa_{2}%
[\mathop{\mathrm{Tr}}(\Phi_{h}^{\dagger}\Phi+\Phi^{\dagger}\Phi_{h}%
)]^{2}-\kappa_{3}\mathop{\mathrm{Tr}}(\Phi_{h}^{\dagger}\Phi+\Phi^{\dagger
}\Phi_{h})\mathop{\mathrm{Tr}}(\Phi_{h}^{\dagger}\Phi_{h})-\kappa
_{4}[\mathop{\mathrm{Tr}}(\Phi_{h}^{\dagger}\Phi_{h})]^{2}\nonumber\\
&  -\xi_{1}\mathop{\mathrm{Tr}}(\Phi_{h}^{\dagger}\Phi\Phi^{\dagger}\Phi
+\Phi_{h}\Phi^{\dagger}\Phi\Phi^{\dagger})-\xi_{2}\mathop{\mathrm{Tr}}(\Phi
_{h}^{\dagger}\Phi\Phi_{h}^{\dagger}\Phi+\Phi^{\dagger}\Phi_{h}\Phi^{\dagger
}\Phi_{h})-\xi_{3}\mathop{\mathrm{Tr}}(\Phi^{\dagger}\Phi_{h}\Phi_{h}%
^{\dagger}\Phi_{h}+\Phi\Phi_{h}^{\dagger}\Phi_{h}\Phi_{h}^{\dagger})-\xi
_{4}\mathop{\mathrm{Tr}}(\Phi_{h}^{\dagger}\Phi_{h})^{2}{\nonumber}\nonumber\\
&  +\mathop{\mathrm{Tr}}(\Phi_{h}^{\dagger}\Phi_{h}E_{1}+\Phi_{h}\Phi
_{h}^{\dagger}E_{1})+c_{1}^{\ast}[(\det\Phi-\det\Phi_{h}^{\dagger})^{2}%
+(\det\Phi^{\dagger}-\det\Phi_{h})^{2}]+c_{1E}^{\ast}(\det\Phi_{h}-\det
\Phi_{h}^{\dagger})^{2}{\nonumber}\nonumber\\
&  +\frac{h_{1}^{\ast}}{2}\mathop{\mathrm{Tr}}(\Phi_{h}^{\dagger}\Phi
+\Phi^{\dagger}\Phi_{h})\mathop{\mathrm{Tr}}(L_{\mu}^{2}+R_{\mu}^{2}%
)+\frac{h_{1E}^{\ast}}{2}\mathop{\mathrm{Tr}}(\Phi_{h}^{\dagger}\Phi
_{h})\mathop{\mathrm{Tr}}(L_{\mu}^{2}+R_{\mu}^{2}){\nonumber}\nonumber\\
&  +h_{2}^{\ast}\mathop{\mathrm{Tr}}(\Phi_{h}^{\dagger}L_{\mu}L^{\mu}\Phi
+\Phi^{\dagger}L_{\mu}L^{\mu}\Phi_{h}+R_{\mu}\Phi_{h}^{\dagger}\Phi R^{\mu
}+R_{\mu}\Phi^{\dagger}\Phi_{h}R^{\mu})+h_{2E}^{\ast}%
\mathop{\mathrm{Tr}} [\vert L_{\mu}\Phi_{h} \vert ^{2}+\vert \Phi_{h} R_{\mu} \vert ^{2}]{\nonumber}\nonumber\\
&  +2h_{3}^{\ast}\mathop{\mathrm{Tr}}(L_{\mu}\Phi_{h}R^{\mu}\Phi^{\dagger
}+L_{\mu}\Phi R^{\mu}\Phi_{h}^{\dagger})+2h_{3E}^{\ast}%
\mathop{\mathrm{Tr}}(L_{\mu}\Phi_{h}R^{\mu}\Phi_{h}^{\dagger})\text{ .}
\end{align}
When fitting the coefficients of this Lagrangian to the data, Ref.~\cite{Parganlija:2016yxq} dropped for simplicity the terms inducing the mixing between light and heavy pseudoscalars. As we have discussed in Sec.~\ref{app:adding-Ph}, in this case, $\kappa_{q}$ invariance also leads to the cancellation of the $\kappa_{q}$-independent pieces in the interaction between ALPs and $P_{h}$. On the other hand, the dropped terms may be essential for the ALPs.

Indeed, let us consider the second term, which induces such mixing, and perform the usual diagonalization~\eqref{eq:diagonalization-supplemental}. Among the other terms, it includes the ALP-meson mixing, which does not vanish in the limit $c_{u,d,s}\to 0$:
\begin{multline}
   \mathcal{L}_{\alpha}  \subset \alpha \frac{f_{\pi}}{f_{a}}c_{G}a(x)\bigg[\eta_{1440}(x) \frac{\sqrt{2}  \left(m_a^2-m_{\pi^{0}}^2\right) \left(m_{\eta}^2-m_{\eta'}^2\right)}{3 \left(m_a^2-m_{\eta}^2\right) \left(m_a^2-m_{\eta'}^2\right)}- \\ -\eta_{1295}(x)\frac{ \left(m_a^2
   \left(m_{\eta'}^2+2 m_{\eta}^2-3 m_{\pi^{0}}^2\right)+2 m_{\pi^{0}}^2 m_{\eta'}^2+m_{\eta}^2 \left(m_{\pi^{0}}^2-3 m_{\eta'}^2\right)\right)}{3 \left(m_a^2-m_{\eta}^2\right) \left(m_a^2-m_{\eta'}^2\right)}\bigg]
\end{multline}
Treating it for the moment as the additional mixing, we see that the $c_{G}$-dependent piece involving the ALP-$P_{h}^{0}$ mixing no longer vanishes (there is no corresponding ``counterterm'' from the rotation~\eqref{eq:P-transformed-supplemental},~\eqref{eq:sigma-transformed-supplemental}), and may sizeably contribute to the ALP decay widths. However, including them would require a complete re-analysis of the whole ELSM fit (as additional mixing between $P_{h}$ and $P$ may be non-negligible), which goes beyond the scope of our study. 

Therefore, we conclude that the robustness of the description of the ALP phenomenology is significantly limited. Nevertheless, incorporating the interactions in \textsc{Mathematica} notebook, we provide a state-of-the-art analysis that can be easily improved in the future once the meson spectroscopy becomes more mature.

\section{Production of ALPs}
\label{app:production}

Below, we briefly discuss our method to describe the production modes of ALPs (more details may be found in Ref.~\cite{Kyselov:2025uez}).

The commonly adopted approach to describe ALP production~\cite{Jerhot:2022chi,Antel:2023hkf} was to approximate the flux of ALPs by the formula
\begin{equation}
    \frac{d^{2}N_{a}}{d\phi_{a}dE_{a}} = \left(\frac{d^{2}N_{a}}{d\phi_{a}dE_{a}}\right)_{\text{direct+light decays}} + \sum_{B,X_{s/d}}N_{B}\cdot \text{Br}(B\to X_{s/d}a)\frac{df^{(B)}_{a}}{d\phi_{a}dE_{a}}
    \label{eq:alp-flux-approximation}
\end{equation}
Here, $\phi,E$ stand for the polar angle and energy correspondingly.
\begin{equation}
\left(\frac{d^{2}N_{a}}{d\phi_{a}dE_{a}}\right)_{\text{direct+light decays}} = \sum_{P^{0}}|\theta_{P^{0}a}|^{2}\frac{d^{2}N_{P^{0}}}{d\phi_{P^{0}}dE_{P^{0}}}\bigg|_{\phi_{P^{0}},E_{P^{0}}\to \phi_{a},E_{a}}
\label{eq:alp-flux-approximation-1}
\end{equation}
describes the contribution of the ALPs produced by deep inelastic scatterings and decays of light mesons, with $\phi_{P^{0}},E_{P^{0}}\to \phi_{a},E_{a}$ being some transformation relating the meson's kinematics to the ALP kinematics. The second summand describes the production of the ALPs by decays of $B$ mesons, with $N_{B}$ being the total number of $B$ mesons of the given type produced in collisions, $\text{Br}(B\to X_{s/d}a)$ the branching ratio of the decay into an ALP and a hadronic state containing an $s/d$ quark, and $f_{a}^{(B)}$ the distribution in polar angle and energy of the ALPs produced in this decay (it is normalized by one).

The term~\eqref{eq:alp-flux-approximation-1} literally means ``the flux of ALPs is the summed fluxes of mother mesons times the squared mixing angle''.

This approximation has two problems. First, it explicitly depends on unphysical $\kappa_{q}$ parameters entering $\theta_{P^{0}a}$. Second, it wrongly describes the mass dependence of the ALP flux, both in terms of the integrated value and the kinematics (the proposed $\phi,E$ transformation is ambiguous and leads to unphysical bumps in the angular distribution of ALPs at low center-of-mass frame collision energy). A clear illustration is to consider the ALPs from the mixing with pions; while $\simeq 60-70\%$ of pions are produced by decays of heavier mesons, this is not the case of the ALPs with $m_{a}\gg m_{\pi^{0}}$.

As we have discussed in the main paper, we explicitly decompose this production into separate modes: the ALP production in the proton bremsstrahlung (the initial state radiation process), quark fragmentation (mostly the final state radiation process relevant for masses $m_{a}\lesssim 2\text{ GeV}$), gluon fusion/Drell-Yan process (the final state radiation relevant for heavier ALPs), and decays of light mesons (see also Ref.~\cite{Kyselov:2024dmi} for the definition of the processes). The descriptions of the proton bremsstrahlung and gluon fusion processes may be found in Refs.~\cite{Blinov:2021say,DallaValleGarcia:2023xhh,Kyselov:2024dmi}, whereas below, we discuss the production in decays of mesons and the quark fragmentation.

\subsection{Decays of mesons}
\label{app:alp-decays-mesons}

Let us now present the method to calculate the branching ratios of various mesons, denoted by $h$, into ALPs. First, using the Lagrangian with the ALP interactions, we compute the matrix element of the process, $\mathcal{M}$. Then, we calculate the decay widths using the standard formulas~\cite{ParticleDataGroup:2024cfk}:
\begin{equation}
    \Gamma^{\text{2-body}}_{h\to a} = \frac{|\mathcal{M}|^{2}}{8\pi} \frac{|\mathbf{p}|}{m_{h}^{2}}
\end{equation}
for 2-body decays, where $|\mathbf{p}|$ is the momentum of the decay product at the rest frame of the decaying $h$, and
\begin{equation}
    \Gamma^{\text{3-body}}_{h\to a} =F_{s}\frac{1}{(2\pi)^{3}8m_{h}}\int dE_{1}dE_{3} |\mathcal{M}|^{2},
\end{equation}
for 3-body decays, where $E_{1,3}$ are the energies of the particles 1,3 in the process $h\to 1+2+3$ at $h$'s rest frame, and $F_{s}$ is the symmetry factor, being $1/n!$ if there are $n$ identical particles in the final state and 1 otherwise. We will consider the following decay processes:
\begin{equation}
    \rho^{\pm}\to \pi^{\pm}a, \quad \eta \to \pi^{+}\pi^{-}a, \quad \eta\to 2\pi^{0}a, \quad K_{S}\to  \pi^{0}a, \quad \omega \to  \pi^{+}\pi^{-}a, \quad \omega\to  a\gamma
\end{equation}
As a cross-check, considering the ``pion''-like ALP, we have checked that together with the production via fragmentation, decays of mesons accumulate $\simeq 90\%$ of the SM neutral pion production flux. The processes that we drop are, e.g., $\Lambda \to pa$ and $K^{\pm}\to a\pi^{\pm}$. The first process has too limited phase space and produces very soft ALPs (as most of the momentum is carried away by the outgoing proton). In their turn, $K^{\pm}$ are long-lived and typically do not have time to decay before interacting with the infrastructure surrounding the collision point (or target for beam dump experiments). Given the complexity of describing its decays and also the suppression of the corresponding event yield, it can be safely neglected.

\subsection{Fragmentation}
\label{app:fragmentation}
Qualitatively, the production in the quark fragmentation effectively replaces a meson $P^{0}$ occurring at the last stage of the fragmentation chain with the ALP $a$. 

To account for the proper ALP mass dependence of the flux, we incorporate the production of ALPs via mixing with $P^{0} = \pi^{0},\eta,\eta'$ in \texttt{PYTHIA8}~\cite{Bierlich:2022pfr}. The production of the ALPs from the mixing with the given meson $P^{0}$ is described by $P^{0}$'s fragmentation function, carefully tuned to the data on the meson fluxes available at different facilities and having at most $\mathcal{O}(1)$ uncertainty. More details can be found in Ref.~\cite{Kyselov:2025uez}.

The mixing with heavier mesons (axial-vector mesons $A$ and heavy pseudoscalars $P^{0}_{h}$) cannot be properly accommodated, as \texttt{PYTHIA8} misses these particles in the spectrum. Even if including them, it would be unrealistic to properly accommodate the contribution of these mesons because of the complexity of properly tuning \texttt{PYTHIA8} setup to describe. Therefore, our estimate of the production in this channel is conservative. 

The rate of the replacement $P^{0}\to a$ cannot be given by the mixing angle $\theta_{P^{0}a}$, as it is clearly $\kappa_{q}$-dependent and misses contributions from the direct operator (see Fig.~\ref{fig:mixing-direct}). Instead, we replace it with the ``effective'' rate $\Theta_{P^{0}a}$, which we define from some typical processes involving the $P^{0}a$ mixing and then extrapolate to the whole fragmentation process. Schematically, this rate has the form
\begin{equation}
    \Theta_{P^{0}a} = \theta_{P^{0}a} + \dots,
\end{equation}
where $\dots$ stands for the direct operator contribution and other mixings.

Specifically, to define $\Theta$ for the mixing with $P^{0}=\pi^{0},\eta,\eta'$, we consider, correspondingly, the following processes:
\begin{equation}
    \pi^{0}\pi^{0}\to \pi^{0}a, \quad \pi^{0}\eta \to \pi^{0}a, \quad \pi^{0}\eta' \to \pi^{0}a
\end{equation}
The specific processes have been chosen because they are the simplest scattering processes and because the $P^{0}a$ mixing directly contributes to them. Namely, their matrix elements have the form 
\begin{equation}
\mathcal{M}_{\pi^{0}P^{0}\to \pi^{0} a } = \Theta_{\pi^{0}a}\cdot \tilde{\mathcal{M}},
\end{equation}
Explicitly, we get
\begin{multline}
\Theta_{\pi^{0}a}= \theta_{\pi^{0}a}+c_G \left[\kappa_u-\kappa_d-\delta  \left(\kappa_d+\kappa_u\right) \left(\sqrt{3} \theta_{\pi^{0}\eta'}+\sqrt{6} \theta_{\pi^{0}\eta}+1\right)\right]\\ -\frac{1}{3} \delta  \left(3 \theta_{\pi^{0}\eta'}+3 \sqrt{2}\theta_{\pi^{0}\eta}+\sqrt{3}\right) \theta_{\eta'a}-\frac{1}{3} \delta \theta_{\eta a} \left(3 \sqrt{2} \theta_{\pi^{0}\eta'}+6 \theta_{\pi^{0}\eta}+\sqrt{6}\right),
\label{eq:theta-eff-pi0a}
\end{multline}
\begin{multline}
\Theta_{\eta a}= \theta_{\eta a} +\frac{1}{2} c_G \left[\kappa_d \left(\delta  \left(2 \sqrt{2} \theta_{\pi\eta'}+\theta_{\pi\eta
   }+\sqrt{6}\right)+\sqrt{6}\right)+\kappa_u \left(\sqrt{6}-\delta  \left(2 \sqrt{2} \theta_{\pi\eta'}+\theta_{\pi\eta
   }+\sqrt{6}\right)\right)\right]\\ -\frac{1}{2} \delta  \theta_{\pi^{0}a} \left(2\sqrt{2}\theta_{\pi\eta'}+\theta_{\pi\eta}+\sqrt{6}\right) +\frac{\theta_{\eta'a}}{\sqrt{2}}
\label{eq:theta-eff-etaa}
\end{multline}
\begin{multline}
    \Theta_{\eta'a} = \theta_{\eta'a} + c_G \left[\kappa_d \left(\delta  \left(-\theta_{\pi^{0}\eta'}+2 \sqrt{2} \theta_{\pi^{0}\eta}+\sqrt{3}\right)+\sqrt{3}\right)+\kappa_u\left(\delta  \left(\theta_{\pi^{0}\eta'}-2 \sqrt{2} \theta_{\pi^{0}\eta}-\sqrt{3}\right)+\sqrt{3}\right)\right] +\\+ \delta \cdot\theta_{\pi^{0}a} \left(\theta_{\pi^{0}\eta'}-2 \sqrt{2} \theta_{\pi^{0}\eta}-\sqrt{3}\right)+\sqrt{2} \theta_{\eta a}
\label{eq:theta-eff-etapra}
\end{multline}
Here,
\begin{equation}
\theta_{\pi^{0}\eta} = -\frac{\sqrt{\frac{2}{3}} m_{\pi^{0}}^2}{m_{\eta}^2-m_{\pi^{0}}^2}, \quad  \theta_{\pi^{0}\eta'} = -\frac{m_{\pi^{0}}^2}{\sqrt{3} \left(m_{\eta'}^2-m_{\pi^{0}}^2\right)}
\label{eq:mixing-ChPT}
\end{equation}
are the mixing angles between $\pi^{0}$ and $\eta/\eta'$ (modulus $\delta$) originating from the pure ChPT (see the mass matrix~\eqref{eq:mass-matrix}).

Inserting them and the ALP-$P^{0}$ mixing angles~\eqref{eq:mixing-pi0-supplemental}-\eqref{eq:mixing-etapr-supplemental} in Eqs.~\eqref{eq:theta-eff-pi0a}-\eqref{eq:theta-eff-etapra}, it can be seen that the $\kappa_{q}$ dependence drops out.

Note that using $\Theta_{P^{0}a}$ instead of $\theta_{P^{0}a}$ is also needed in the case of ALPs with the dominant coupling to quarks.

\bibliography{bib.bib}

@article{Peccei:1977hh,
    author = "Peccei, R. D. and Quinn, Helen R.",
    title = "{CP Conservation in the Presence of Instantons}",
    reportNumber = "ITP-568-STANFORD",
    doi = "10.1103/PhysRevLett.38.1440",
    journal = "Phys. Rev. Lett.",
    volume = "38",
    pages = "1440--1443",
    year = "1977"
}

@article{Belle-II:2023ueh,
    author = "Adachi, I. and others",
    collaboration = "Belle-II",
    title = "{Search for a long-lived spin-0 mediator in b\textrightarrow{}s transitions at the Belle II experiment}",
    eprint = "2306.02830",
    archivePrefix = "arXiv",
    primaryClass = "hep-ex",
    reportNumber = "Belle II Preprint 2023-009, KEK Preprint 2023-7",
    doi = "10.1103/PhysRevD.108.L111104",
    journal = "Phys. Rev. D",
    volume = "108",
    number = "11",
    pages = "L111104",
    year = "2023"
}

@article{Giacosa:2024epf,
    author = "Giacosa, Francesco and Kov\'acs, P\'eter and Jafarzade, Shahriyar",
    title = "{Ordinary and exotic mesons in the extended Linear Sigma Model}",
    eprint = "2407.18348",
    archivePrefix = "arXiv",
    primaryClass = "hep-ph",
    month = "7",
    year = "2024"
}

@article{Georgi:1986df,
    author = "Georgi, Howard and Kaplan, David B. and Randall, Lisa",
    title = "{Manifesting the Invisible Axion at Low-energies}",
    reportNumber = "HUTP-86/A004",
    doi = "10.1016/0370-2693(86)90688-X",
    journal = "Phys. Lett. B",
    volume = "169",
    pages = "73--78",
    year = "1986"
}

@article{Kyselov:2025uez,
    author = "Kyselov, Yehor and Mrenna, Stephen and Ovchynnikov, Maksym",
    title = "{New physics particles mixing with mesons: production in the fragmentation chain}",
    eprint = "2504.06828",
    archivePrefix = "arXiv",
    primaryClass = "hep-ph",
    reportNumber = "CERN-TH-2025-073, FERMILAB-PUB-25-0200-CSAID",
    month = "4",
    year = "2025"
}

@article{SHiP:2025ows,
    author = "Albanese, R. and others",
    collaboration = "SHiP, HI-ECN3 Project Team",
    title = "{SHiP experiment at the SPS Beam Dump Facility}",
    eprint = "2504.06692",
    archivePrefix = "arXiv",
    primaryClass = "hep-ex",
    month = "4",
    year = "2025"
}

@article{Witten:1980sp,
    author = "Witten, Edward",
    title = "{Large N Chiral Dynamics}",
    reportNumber = "HUTP-80/A005",
    doi = "10.1016/0003-4916(80)90325-5",
    journal = "Annals Phys.",
    volume = "128",
    pages = "363",
    year = "1980"
}

@article{Alves:2017avw,
    author = "Alves, Daniele S. M. and Weiner, Neal",
    title = "{A viable QCD axion in the MeV mass range}",
    eprint = "1710.03764",
    archivePrefix = "arXiv",
    primaryClass = "hep-ph",
    reportNumber = "LA-UR-17-29295",
    doi = "10.1007/JHEP07(2018)092",
    journal = "JHEP",
    volume = "07",
    pages = "092",
    year = "2018"
}

@article{Bardeen:1986yb,
    author = "Bardeen, William A. and Peccei, R. D. and Yanagida, T.",
    title = "{CONSTRAINTS ON VARIANT AXION MODELS}",
    reportNumber = "DESY-86-054, MPI-PAE/PTh 27/86",
    doi = "10.1016/0550-3213(87)90003-4",
    journal = "Nucl. Phys. B",
    volume = "279",
    pages = "401--428",
    year = "1987"
}

@article{Herrera-Siklody:1996tqr,
    author = "Herrera-Siklody, P. and Latorre, J. I. and Pascual, P. and Taron, J.",
    title = "{Chiral effective Lagrangian in the large N(c) limit: The Nonet case}",
    eprint = "hep-ph/9610549",
    archivePrefix = "arXiv",
    reportNumber = "UB-ECM-PF-96-16",
    doi = "10.1016/S0550-3213(97)00260-5",
    journal = "Nucl. Phys. B",
    volume = "497",
    pages = "345--386",
    year = "1997"
}

@article{Kaiser:2000gs,
    author = "Kaiser, Roland and Leutwyler, H.",
    title = "{Large N(c) in chiral perturbation theory}",
    eprint = "hep-ph/0007101",
    archivePrefix = "arXiv",
    reportNumber = "BUTP-00-19",
    doi = "10.1007/s100520000499",
    journal = "Eur. Phys. J. C",
    volume = "17",
    pages = "623--649",
    year = "2000"
}

@article{Parganlija:2016yxq,
    author = "Parganlija, Denis and Giacosa, Francesco",
    title = "{Excited Scalar and Pseudoscalar Mesons in the Extended Linear Sigma Model}",
    eprint = "1612.09218",
    archivePrefix = "arXiv",
    primaryClass = "hep-ph",
    doi = "10.1140/epjc/s10052-017-4962-y",
    journal = "Eur. Phys. J. C",
    volume = "77",
    number = "7",
    pages = "450",
    year = "2017"
}

@article{Parganlija:2012fy,
    author = "Parganlija, Denis and Kovacs, Peter and Wolf, Gyorgy and Giacosa, Francesco and Rischke, Dirk H.",
    title = "{Meson vacuum phenomenology in a three-flavor linear sigma model with (axial-)vector mesons}",
    eprint = "1208.0585",
    archivePrefix = "arXiv",
    primaryClass = "hep-ph",
    doi = "10.1103/PhysRevD.87.014011",
    journal = "Phys. Rev. D",
    volume = "87",
    number = "1",
    pages = "014011",
    year = "2013"
}

@article{ParticleDataGroup:2024cfk,
    author = "Navas, S. and others",
    collaboration = "Particle Data Group",
    title = "{Review of particle physics}",
    doi = "10.1103/PhysRevD.110.030001",
    journal = "Phys. Rev. D",
    volume = "110",
    number = "3",
    pages = "030001",
    year = "2024"
}

@article{Cornella:2023kjq,
    author = "Cornella, Claudia and Galda, Anne Mareike and Neubert, Matthias and Wyler, Daniel",
    title = "{$K^\pm\to\pi^\pm a$~at next-to-leading order in chiral perturbation theory and updated bounds on ALP couplings}",
    eprint = "2308.16903",
    archivePrefix = "arXiv",
    primaryClass = "hep-ph",
    doi = "10.1007/JHEP06(2024)029",
    journal = "JHEP",
    volume = "06",
    pages = "029",
    year = "2024"
}

@techreport{AlemanyFernandez:2927631,
      author        = "Alemany Fernandez, Reyes et al.",
      title         = "{Summary Report of the Physics Beyond Colliders Study at
                       CERN}",
      institution   = "CERN",
      reportNumber  = "CERN-PBC-REPORT-2025-003",
      address       = "Geneva",
      year          = "2025",
      url           = "https://cds.cern.ch/record/2927631",
}

@article{DallaValleGarcia:2023xhh,
    author = "Dalla Valle Garcia, Giovani and Kahlhoefer, Felix and Ovchynnikov, Maksym and Zaporozhchenko, Andrii",
    title = "{Phenomenology of axionlike particles with universal fermion couplings revisited}",
    eprint = "2310.03524",
    archivePrefix = "arXiv",
    primaryClass = "hep-ph",
    reportNumber = "TTP23-042, P3H-23-070",
    doi = "10.1103/PhysRevD.109.055042",
    journal = "Phys. Rev. D",
    volume = "109",
    number = "5",
    pages = "055042",
    year = "2024"
}

@article{Foroughi-Abari:2021zbm,
    author = "Foroughi-Abari, Saeid and Ritz, Adam",
    title = "{Dark sector production via proton bremsstrahlung}",
    eprint = "2108.05900",
    archivePrefix = "arXiv",
    primaryClass = "hep-ph",
    doi = "10.1103/PhysRevD.105.095045",
    journal = "Phys. Rev. D",
    volume = "105",
    number = "9",
    pages = "095045",
    year = "2022"
}

@article{Black:1998wt,
    author = "Black, Deirdre and Fariborz, Amir H. and Sannino, Francesco and Schechter, Joseph",
    title = "{Putative light scalar nonet}",
    eprint = "hep-ph/9808415",
    archivePrefix = "arXiv",
    reportNumber = "YCTP-21-98, SU-4240-683",
    doi = "10.1103/PhysRevD.59.074026",
    journal = "Phys. Rev. D",
    volume = "59",
    pages = "074026",
    year = "1999"
}

@article{Foroughi-Abari:2024xlj,
    author = "Foroughi-Abari, Saeid and Reimitz, Peter and Ritz, Adam",
    title = "{A Closer Look at Dark Vector Splitting Functions in Proton Bremsstrahlung}",
    eprint = "2409.09123",
    archivePrefix = "arXiv",
    primaryClass = "hep-ph",
    month = "9",
    year = "2024"
}

@article{Blinov:2021say,
    author = "Blinov, Nikita and Kowalczyk, Elizabeth and Wynne, Margaret",
    title = "{Axion-like particle searches at DarkQuest}",
    eprint = "2112.09814",
    archivePrefix = "arXiv",
    primaryClass = "hep-ph",
    reportNumber = "FERMILAB-PUB-21-749-V",
    doi = "10.1007/JHEP02(2022)036",
    journal = "JHEP",
    volume = "02",
    pages = "036",
    year = "2022"
}

@article{Altarelli:1977zs,
    author = "Altarelli, Guido and Parisi, G.",
    title = "{Asymptotic Freedom in Parton Language}",
    reportNumber = "LPTENS-77-6",
    doi = "10.1016/0550-3213(77)90384-4",
    journal = "Nucl. Phys. B",
    volume = "126",
    pages = "298--318",
    year = "1977"
}

@article{Gorkavenko:2023nbk,
    author = "Gorkavenko, Volodymyr and Jashal, Brij Kishor and Kholoimov, Valerii and Kyselov, Yehor and Mendoza, Diego and Ovchynnikov, Maksym and Oyanguren, Arantza and Svintozelskyi, Volodymyr and Zhuo, Jiahui",
    title = "{LHCb potential to discover long-lived new physics particles with lifetimes above 100~ps}",
    eprint = "2312.14016",
    archivePrefix = "arXiv",
    primaryClass = "hep-ph",
    doi = "10.1140/epjc/s10052-024-12906-3",
    journal = "Eur. Phys. J. C",
    volume = "84",
    number = "6",
    pages = "608",
    year = "2024"
}

@article{NA62:2025yzs,
    author = "Cortina Gil, Eduardo and others",
    collaboration = "NA62",
    title = "{Search for hadronic decays of feebly-interacting particles at NA62}",
    eprint = "2502.04241",
    archivePrefix = "arXiv",
    primaryClass = "hep-ex",
    reportNumber = "CERN-EP-2025-012",
    month = "2",
    year = "2025"
}

@article{Fujiwara:1984mp,
    author = "Fujiwara, Takanori and Kugo, Taichiro and Terao, Haruhiko and Uehara, Shozo and Yamawaki, Koichi",
    title = "{Nonabelian Anomaly and Vector Mesons as Dynamical Gauge Bosons of Hidden Local Symmetries}",
    reportNumber = "KUNS-764",
    doi = "10.1143/PTP.73.926",
    journal = "Prog. Theor. Phys.",
    volume = "73",
    pages = "926",
    year = "1985"
}

@article{Bierlich:2022pfr,
    author = "Bierlich, Christian and others",
    title = "{A comprehensive guide to the physics and usage of PYTHIA 8.3}",
    eprint = "2203.11601",
    archivePrefix = "arXiv",
    primaryClass = "hep-ph",
    reportNumber = "LU-TP 22-16, MCNET-22-04, FERMILAB-PUB-22-227-SCD",
    doi = "10.21468/SciPostPhysCodeb.8",
    journal = "SciPost Phys. Codeb.",
    volume = "2022",
    pages = "8",
    year = "2022"
}

@article{Bauer:2021wjo,
    author = "Bauer, Martin and Neubert, Matthias and Renner, Sophie and Schnubel, Marvin and Thamm, Andrea",
    title = "{Consistent Treatment of Axions in the Weak Chiral Lagrangian}",
    eprint = "2102.13112",
    archivePrefix = "arXiv",
    primaryClass = "hep-ph",
    reportNumber = "IPPP/20-82, MITP/21-007, ZU-TH-01/21",
    doi = "10.1103/PhysRevLett.127.081803",
    journal = "Phys. Rev. Lett.",
    volume = "127",
    number = "8",
    pages = "081803",
    year = "2021"
}

@article{Cirigliano:2011ny,
    author = "Cirigliano, Vincenzo and Ecker, Gerhard and Neufeld, Helmut and Pich, Antonio and Portoles, Jorge",
    title = "{Kaon Decays in the Standard Model}",
    eprint = "1107.6001",
    archivePrefix = "arXiv",
    primaryClass = "hep-ph",
    reportNumber = "FTUV-11-0729, IFIC-11-02, UWTHPH-2011-25",
    doi = "10.1103/RevModPhys.84.399",
    journal = "Rev. Mod. Phys.",
    volume = "84",
    pages = "399",
    year = "2012"
}

@article{Fitzpatrick:2023xks,
    author = "Fitzpatrick, Patrick J. and Hochberg, Yonit and Kuflik, Eric and Ovadia, Rotem and Soreq, Yotam",
    title = "{Dark matter through the axion-gluon portal}",
    eprint = "2306.03128",
    archivePrefix = "arXiv",
    primaryClass = "hep-ph",
    doi = "10.1103/PhysRevD.108.075003",
    journal = "Phys. Rev. D",
    volume = "108",
    number = "7",
    pages = "075003",
    year = "2023"
}

@article{Hochberg:2018rjs,
    author = "Hochberg, Yonit and Kuflik, Eric and Mcgehee, Robert and Murayama, Hitoshi and Schutz, Katelin",
    title = "{Strongly interacting massive particles through the axion portal}",
    eprint = "1806.10139",
    archivePrefix = "arXiv",
    primaryClass = "hep-ph",
    reportNumber = "DESY-18-101, IPMU18-0114",
    doi = "10.1103/PhysRevD.98.115031",
    journal = "Phys. Rev. D",
    volume = "98",
    number = "11",
    pages = "115031",
    year = "2018"
}

@article{Dolan:2014ska,
    author = "Dolan, Matthew J. and Kahlhoefer, Felix and McCabe, Christopher and Schmidt-Hoberg, Kai",
    title = "{A taste of dark matter: Flavour constraints on pseudoscalar mediators}",
    eprint = "1412.5174",
    archivePrefix = "arXiv",
    primaryClass = "hep-ph",
    reportNumber = "DESY-14-238, SLAC-PUB-16179",
    doi = "10.1007/JHEP03(2015)171",
    journal = "JHEP",
    volume = "03",
    pages = "171",
    year = "2015",
    note = "[Erratum: JHEP 07, 103 (2015)]"
}

@article{Nomura:2008ru,
    author = "Nomura, Yasunori and Thaler, Jesse",
    title = "{Dark Matter through the Axion Portal}",
    eprint = "0810.5397",
    archivePrefix = "arXiv",
    primaryClass = "hep-ph",
    doi = "10.1103/PhysRevD.79.075008",
    journal = "Phys. Rev. D",
    volume = "79",
    pages = "075008",
    year = "2009"
}

@article{Irastorza:2018dyq,
    author = "Irastorza, Igor G. and Redondo, Javier",
    title = "{New experimental approaches in the search for axion-like particles}",
    eprint = "1801.08127",
    archivePrefix = "arXiv",
    primaryClass = "hep-ph",
    doi = "10.1016/j.ppnp.2018.05.003",
    journal = "Prog. Part. Nucl. Phys.",
    volume = "102",
    pages = "89--159",
    year = "2018"
}

@article{Graham:2015ouw,
    author = "Graham, Peter W. and Irastorza, Igor G. and Lamoreaux, Steven K. and Lindner, Axel and van Bibber, Karl A.",
    title = "{Experimental Searches for the Axion and Axion-Like Particles}",
    eprint = "1602.00039",
    archivePrefix = "arXiv",
    primaryClass = "hep-ex",
    doi = "10.1146/annurev-nucl-102014-022120",
    journal = "Ann. Rev. Nucl. Part. Sci.",
    volume = "65",
    pages = "485--514",
    year = "2015"
}

@article{Marsh:2015xka,
    author = "Marsh, David J. E.",
    title = "{Axion Cosmology}",
    eprint = "1510.07633",
    archivePrefix = "arXiv",
    primaryClass = "astro-ph.CO",
    reportNumber = "KCL-PH-TH-2015-50",
    doi = "10.1016/j.physrep.2016.06.005",
    journal = "Phys. Rept.",
    volume = "643",
    pages = "1--79",
    year = "2016"
}

@inproceedings{Essig:2013lka,
    author = "Essig, Rouven and others",
    title = "{Working Group Report: New Light Weakly Coupled Particles}",
    booktitle = "{Snowmass 2013}: {Snowmass on the Mississippi}",
    eprint = "1311.0029",
    archivePrefix = "arXiv",
    primaryClass = "hep-ph",
    reportNumber = "YITP-SB-36, FERMILAB-CONF-13-653",
    month = "10",
    year = "2013"
}

@article{NA62:2023qyn,
    author = "Cortina Gil, Eduardo and others",
    collaboration = "NA62",
    title = "{Search for dark photon decays to $\mu^{+}\mu^{-}$ at NA62}",
    eprint = "2303.08666",
    archivePrefix = "arXiv",
    primaryClass = "hep-ex",
    reportNumber = "CERN-EP-2023-032",
    doi = "10.1007/JHEP09(2023)035",
    journal = "JHEP",
    volume = "09",
    pages = "035",
    year = "2023"
}

@article{Afik:2023mhj,
    author = {Afik, Yoav and D\"obrich, Babette and Jerhot, Jan and Soreq, Yotam and Tobioka, Kohsaku},
    title = "{Probing long-lived axions at the KOTO experiment}",
    eprint = "2303.01521",
    archivePrefix = "arXiv",
    primaryClass = "hep-ph",
    reportNumber = "IRMP-CP3-23-11, IRMP-CP3-23-10, MPP-2023-40, KEK-TH-2499",
    doi = "10.1103/PhysRevD.108.055007",
    journal = "Phys. Rev. D",
    volume = "108",
    number = "5",
    pages = "055007",
    year = "2023"
}

@article{Chakraborty:2021wda,
    author = "Chakraborty, Sabyasachi and Kraus, Manfred and Loladze, Vazha and Okui, Takemichi and Tobioka, Kohsaku",
    title = "{Heavy QCD axion in b\textrightarrow{}s transition: Enhanced limits and projections}",
    eprint = "2102.04474",
    archivePrefix = "arXiv",
    primaryClass = "hep-ph",
    reportNumber = "KEK-TH-2295",
    doi = "10.1103/PhysRevD.104.055036",
    journal = "Phys. Rev. D",
    volume = "104",
    number = "5",
    pages = "055036",
    year = "2021"
}

@article{Mikulenko:2023olf,
    author = "Mikulenko, Oleksii and Bondarenko, Kyrylo and Boyarsky, Alexey and Ruchayskiy, Oleg",
    title = "{New physics at the Intensity Frontier: how much can we learn and how?}",
    eprint = "2312.00659",
    archivePrefix = "arXiv",
    primaryClass = "hep-ph",
    month = "12",
    year = "2023"
}

@article{Kyselov:2024dmi,
    author = "Kyselov, Yehor and Ovchynnikov, Maksym",
    title = "{Searches for long-lived dark photons at proton accelerator experiments}",
    eprint = "2409.11096",
    archivePrefix = "arXiv",
    primaryClass = "hep-ph",
    reportNumber = "CERN-TH-2024-157",
    doi = "10.1103/PhysRevD.111.015030",
    journal = "Phys. Rev. D",
    volume = "111",
    number = "1",
    pages = "015030",
    year = "2025"
}

@article{Gori:2020xvq,
    author = "Gori, Stefania and Perez, Gilad and Tobioka, Kohsaku",
    title = "{KOTO vs. NA62 Dark Scalar Searches}",
    eprint = "2005.05170",
    archivePrefix = "arXiv",
    primaryClass = "hep-ph",
    doi = "10.1007/JHEP08(2020)110",
    journal = "JHEP",
    volume = "08",
    pages = "110",
    year = "2020"
}

@article{Bai:2024lpq,
    author = "Bai, Yang and Chen, Ting-Kuo and Liu, Jia and Ma, Xiaolin",
    title = "{Wess-Zumino-Witten Interactions of Axions}",
    eprint = "2406.11948",
    archivePrefix = "arXiv",
    primaryClass = "hep-ph",
    month = "6",
    year = "2024"
}

@article{Bauer:2020jbp,
    author = "Bauer, Martin and Neubert, Matthias and Renner, Sophie and Schnubel, Marvin and Thamm, Andrea",
    title = "{The Low-Energy Effective Theory of Axions and ALPs}",
    eprint = "2012.12272",
    archivePrefix = "arXiv",
    primaryClass = "hep-ph",
    reportNumber = "IPPP/20/69, MITP/20-070 SISSA 30/2020/FISI, ZH-TH-47/20",
    doi = "10.1007/JHEP04(2021)063",
    journal = "JHEP",
    volume = "04",
    pages = "063",
    year = "2021"
}

@article{CHARM:1985anb,
    author = "Bergsma, F. and others",
    collaboration = "CHARM",
    title = "{Search for Axion Like Particle Production in 400-{GeV} Proton - Copper Interactions}",
    reportNumber = "CERN-EP-85-38",
    doi = "10.1016/0370-2693(85)90400-9",
    journal = "Phys. Lett. B",
    volume = "157",
    pages = "458--462",
    year = "1985"
}

@article{Wilczek:1977pj,
    author = "Wilczek, Frank",
    title = "{Problem of Strong  $P$  and  $T$  Invariance in the Presence of Instantons}",
    reportNumber = "Print-77-0939 (COLUMBIA)",
    doi = "10.1103/PhysRevLett.40.279",
    journal = "Phys. Rev. Lett.",
    volume = "40",
    pages = "279--282",
    year = "1978"
}

@article{Weinberg:1977ma,
    author = "Weinberg, Steven",
    title = "{A New Light Boson?}",
    reportNumber = "HUTP-77/A074",
    doi = "10.1103/PhysRevLett.40.223",
    journal = "Phys. Rev. Lett.",
    volume = "40",
    pages = "223--226",
    year = "1978"
}

@article{Boiarska:2019jym,
    author = "Boiarska, Iryna and Bondarenko, Kyrylo and Boyarsky, Alexey and Gorkavenko, Volodymyr and Ovchynnikov, Maksym and Sokolenko, Anastasia",
    title = "{Phenomenology of GeV-scale scalar portal}",
    eprint = "1904.10447",
    archivePrefix = "arXiv",
    primaryClass = "hep-ph",
    doi = "10.1007/JHEP11(2019)162",
    journal = "JHEP",
    volume = "11",
    pages = "162",
    year = "2019"
}

@article{Cornella:2019uxs,
    author = "Cornella, Claudia and Paradisi, Paride and Sumensari, Olcyr",
    title = "{Hunting for ALPs with Lepton Flavor Violation}",
    eprint = "1911.06279",
    archivePrefix = "arXiv",
    primaryClass = "hep-ph",
    reportNumber = "ZU-TH 46/19",
    doi = "10.1007/JHEP01(2020)158",
    journal = "JHEP",
    volume = "01",
    pages = "158",
    year = "2020"
}

@article{Blumlein:1990ay,
    author = "Blumlein, J. and others",
    title = "{Limits on neutral light scalar and pseudoscalar particles in a proton beam dump experiment}",
    reportNumber = "PHE-90-03",
    doi = "10.1007/BF01548556",
    journal = "Z. Phys. C",
    volume = "51",
    pages = "341--350",
    year = "1991"
}

@article{Bauer:2021mvw,
    author = "Bauer, Martin and Neubert, Matthias and Renner, Sophie and Schnubel, Marvin and Thamm, Andrea",
    title = "{Flavor probes of axion-like particles}",
    eprint = "2110.10698",
    archivePrefix = "arXiv",
    primaryClass = "hep-ph",
    reportNumber = "MITP/21-025, CERN-TH-2021-148, IPPP/21/37",
    doi = "10.1007/JHEP09(2022)056",
    journal = "JHEP",
    volume = "09",
    pages = "056",
    year = "2022"
}

@article{Fariborz:1999gr,
    author = "Fariborz, Amir H. and Schechter, Joseph",
    title = "{Eta-prime ---\ensuremath{>} eta pi pi decay as a probe of a possible lowest lying scalar nonet}",
    eprint = "hep-ph/9902238",
    archivePrefix = "arXiv",
    reportNumber = "SU-4240-693",
    doi = "10.1103/PhysRevD.60.034002",
    journal = "Phys. Rev. D",
    volume = "60",
    pages = "034002",
    year = "1999"
}

@article{Cheng:2021kjg,
    author = "Cheng, Hsin-Chia and Li, Lingfeng and Salvioni, Ennio",
    title = "{A theory of dark pions}",
    eprint = "2110.10691",
    archivePrefix = "arXiv",
    primaryClass = "hep-ph",
    reportNumber = "CERN-TH-2021-150",
    doi = "10.1007/JHEP01(2022)122",
    journal = "JHEP",
    volume = "01",
    pages = "122",
    year = "2022"
}

@article{Guo:2011ir,
    author = "Guo, Feng-Kun and Kubis, Bastian and Wirzba, Andreas",
    title = "{Anomalous decays of eta' and eta into four pions}",
    eprint = "1111.5949",
    archivePrefix = "arXiv",
    primaryClass = "hep-ph",
    doi = "10.1103/PhysRevD.85.014014",
    journal = "Phys. Rev. D",
    volume = "85",
    pages = "014014",
    year = "2012"
}

@article{Aloni:2018vki,
    author = "Aloni, Daniel and Soreq, Yotam and Williams, Mike",
    title = "{Coupling QCD-Scale Axionlike Particles to Gluons}",
    eprint = "1811.03474",
    archivePrefix = "arXiv",
    primaryClass = "hep-ph",
    reportNumber = "CERN-TH-2018-237, MIT-CTP/5080, MIT-CTP-5080",
    doi = "10.1103/PhysRevLett.123.031803",
    journal = "Phys. Rev. Lett.",
    volume = "123",
    number = "3",
    pages = "031803",
    year = "2019"
}

@article{Ovchynnikov:2023cry,
    author = "Ovchynnikov, Maksym and Tastet, Jean-Loup and Mikulenko, Oleksii and Bondarenko, Kyrylo",
    title = "{Sensitivities to feebly interacting particles: Public and unified calculations}",
    eprint = "2305.13383",
    archivePrefix = "arXiv",
    primaryClass = "hep-ph",
    doi = "10.1103/PhysRevD.108.075028",
    journal = "Phys. Rev. D",
    volume = "108",
    number = "7",
    pages = "075028",
    year = "2023"
}

@article{Antel:2023hkf,
    author = "Antel, C. and others",
    title = "{Feebly-interacting particles: FIPs 2022 Workshop Report}",
    eprint = "2305.01715",
    archivePrefix = "arXiv",
    primaryClass = "hep-ph",
    reportNumber = "CERN-TH-2023-061, DESY-23-050, FERMILAB-PUB-23-149-PPD, INFN-23-14-LNF, JLAB-PHY-23-3789, LA-UR-23-21432, MITP-23-015",
    doi = "10.1140/epjc/s10052-023-12168-5",
    journal = "Eur. Phys. J. C",
    volume = "83",
    number = "12",
    pages = "1122",
    year = "2023"
}

@article{Kling:2021fwx,
    author = "Kling, Felix and Trojanowski, Sebastian",
    title = "{Forward experiment sensitivity estimator for the LHC and future hadron colliders}",
    eprint = "2105.07077",
    archivePrefix = "arXiv",
    primaryClass = "hep-ph",
    doi = "10.1103/PhysRevD.104.035012",
    journal = "Phys. Rev. D",
    volume = "104",
    number = "3",
    pages = "035012",
    year = "2021"
}

@misc{Mathematica,
  author = {{Wolfram Research, Inc.}},
  title = {Mathematica, {V}ersion 13.2},
  url = {https://www.wolfram.com/mathematica},
  note = {Champaign, IL, 2022}
}

@article{Batell:2020vqn,
    author = "Batell, Brian and Evans, Jared A. and Gori, Stefania and Rai, Mudit",
    title = "{Dark Scalars and Heavy Neutral Leptons at DarkQuest}",
    eprint = "2008.08108",
    archivePrefix = "arXiv",
    primaryClass = "hep-ph",
    doi = "10.1007/JHEP05(2021)049",
    journal = "JHEP",
    volume = "05",
    pages = "049",
    year = "2021"
}

@article{Beacham:2019nyx,
    author = "Beacham, J. and others",
    title = "{Physics Beyond Colliders at CERN: Beyond the Standard Model Working Group Report}",
    eprint = "1901.09966",
    archivePrefix = "arXiv",
    primaryClass = "hep-ex",
    reportNumber = "CERN-PBC-REPORT-2018-007",
    doi = "10.1088/1361-6471/ab4cd2",
    journal = "J. Phys. G",
    volume = "47",
    number = "1",
    pages = "010501",
    year = "2020"
}

@techreport{Aberle:2839677,
      author        = "Aberle, O and others",
      collaboration = "SHiP",
      title         = "{BDF/SHiP at the ECN3 high-intensity beam facility}",
      institution   = "CERN",
      reportNumber  = "CERN-SPSC-2022-032, SPSC-I-258",
      address       = "Geneva",
      year          = "2022",
      url           = "http://cds.cern.ch/record/2839677",
}

@article{BEBCWA66:1986err,
    author = "Grassler, H. and others",
    collaboration = "BEBC WA66",
    title = "{Prompt Neutrino Production in 400-{GeV} Proton Copper Interactions}",
    reportNumber = "USIP-86-02",
    doi = "10.1016/0550-3213(86)90246-4",
    journal = "Nucl. Phys. B",
    volume = "273",
    pages = "253--274",
    year = "1986"
}

@article{Aielli:2019ivi,
    author = "Aielli, Giulio and others",
    title = "{Expression of interest for the CODEX-b detector}",
    eprint = "1911.00481",
    archivePrefix = "arXiv",
    primaryClass = "hep-ex",
    doi = "10.1140/epjc/s10052-020-08711-3",
    journal = "Eur. Phys. J. C",
    volume = "80",
    number = "12",
    pages = "1177",
    year = "2020"
}

@article{Ilten:2018crw,
    author = "Ilten, Philip and Soreq, Yotam and Williams, Mike and Xue, Wei",
    title = "{Serendipity in dark photon searches}",
    eprint = "1801.04847",
    archivePrefix = "arXiv",
    primaryClass = "hep-ph",
    reportNumber = "MIT-CTP/4976, CERN-TH-2017-282, MIT-CTP-4976",
    doi = "10.1007/JHEP06(2018)004",
    journal = "JHEP",
    volume = "06",
    pages = "004",
    year = "2018"
}

@article{Jerhot:2022chi,
    author = {Jerhot, Jan and D\"obrich, Babette and Ertas, Fatih and Kahlhoefer, Felix and Spadaro, Tommaso},
    title = "{ALPINIST: Axion-Like Particles In Numerous Interactions Simulated and Tabulated}",
    eprint = "2201.05170",
    archivePrefix = "arXiv",
    primaryClass = "hep-ph",
    reportNumber = "TTK-22-04, CP3-22-02",
    doi = "10.1007/JHEP07(2022)094",
    journal = "JHEP",
    volume = "07",
    pages = "094",
    year = "2022"
}

@article{Bauer:2017ris,
    author = "Bauer, Martin and Neubert, Matthias and Thamm, Andrea",
    title = "{Collider Probes of Axion-Like Particles}",
    eprint = "1708.00443",
    archivePrefix = "arXiv",
    primaryClass = "hep-ph",
    reportNumber = "MITP-17-047",
    doi = "10.1007/JHEP12(2017)044",
    journal = "JHEP",
    volume = "12",
    pages = "044",
    year = "2017"
}

\end{document}